\documentclass[preprint,12pt]{elsarticle}

\usepackage{breakurl}             
\usepackage{underscore} 
\usepackage{subfigure}
\usepackage{amsfonts}
\usepackage{amsthm}
\makeatletter
\def\els@aparagraph[#1]#2{\elsparagraph[#1]{#2\@addpunct{.}}}
\def\els@bparagraph#1{\elsparagraph*{#1\@addpunct{.}}}
\makeatother
\usepackage{mathtools}
\usepackage{latexsym}
\usepackage{amssymb}
\newcommand{\remove}[1]{} 
\usepackage[subfigure]{tocloft}
\usepackage{amsfonts}
\usepackage{graphicx}
\usepackage{cleveref}
\usepackage{soul}
\usepackage{amssymb}
\usepackage[dvipsnames]{xcolor}

\newtheorem{definition}{Definition}
\newtheorem{lemma}{Lemma}
\newtheorem{theorem}{Theorem}

\newtheorem{proposition}{Proposition}
\newtheorem{corollary}{Corollary}
\usepackage{amssymb}

\newcommand{\bond}{\!-\!}

\newcommand{\connected}{\mathsf{con}}
\newcommand{\paths}{\mathsf{path}}
\newcommand{\es}{\emptyset}
\newcommand{\state}[2]{\langle {#1}, {#2}\rangle}
\newcommand{\cstate}[3]{\langle {#1}, {#2}, {#3}\rangle}
\newcommand{\trans}[1]{\ensuremath{\stackrel{#1}{\longrightarrow}}}
\newcommand{\guard}[1]{\mathsf{pre}(#1)}
\newcommand{\effects}[1]{\mathsf{post}(#1)}

\newcommand{\effect}[1]{\mathsf{eff}(#1)}

\newcommand{\lastp}[1]{\mathsf{last}_P(#1)}
\newcommand{\lastt}[1]{\mathsf{last}_T(#1)}

\newcommand{\PN}{reversing Petri net } 
\newcommand{\RPN}{\textsc{RPN\ }} 
\newcommand{\btrans}[1]{\ensuremath{\stackrel{#1}{\rightsquigarrow}_{b}}}
\newcommand{\ctrans}[1]{\ensuremath{\stackrel{#1}{\rightsquigarrow}_{c}}}
\newcommand{\otrans}[1]{\ensuremath{\stackrel{#1}{\rightsquigarrow}_{o}}}
\newcommand{\frtrans}[1]{\ensuremath{\stackrel{#1}{\longmapsto}}}
\newcommand{\fctrans}[1]{\ensuremath{\stackrel{#1}{\longmapsto}_{c}}}
\newcommand{\fbtrans}[1]{\ensuremath{\stackrel{#1}{\longmapsto}_{b}}}
\newcommand{\fotrans}[1]{\ensuremath{\stackrel{#1}{\longmapsto}_{o}}}
\newcommand{\comment}[1]{}
\newcommand{\proofend}{\hspace*{\fill} $\Box$}

\newcommand{\rtrans}[1]{\ensuremath{\stackrel{#1}{\rightsquigarrow}}}

\journal{Journal of Logical and Algebraic Methods in Programming}

\begin{document}

\begin{frontmatter}



\title{Reversible Computation in Cyclic Petri Nets}


\author{Anna Philippou}
 \ead{annap@ucy.ac.cy}
\author{Kyriaki Psara}
 \ead{psara.kyriaki@ucy.ac.cy}
 \address{Department of Computer Science\\
 	University of Cyprus\\ Nicosia, Cyprus}

\begin{abstract}
Petri nets are a mathematical language for modeling and reasoning about distributed systems.
In this paper we propose an approach to Petri nets for embedding reversibility, i.e., the ability of reversing
an executed sequence of operations at any point during operation.
Specifically, we introduce machinery and associated semantics to support the three main forms of 
reversibility namely, backtracking, causal reversing, and out-of-causal-order reversing in a variation of
cyclic Petri nets where tokens are persistent and are distinguished from each other by an identity. 
Our formalism is influenced 
by applications in biochemistry but the 
methodology can be applied to a wide range of problems that feature reversibility. In particular, we demonstrate the 
applicability of our approach
with a model of the ERK signalling pathway, 
an example that inherently features reversible behavior.

\end{abstract}

\begin{keyword}
	Reversible computation \sep Petri nets \sep cycles \sep causal reversibility \sep out-of-causal reversibility 

\end{keyword}

\end{frontmatter}

\pagestyle{plain}
\pagenumbering{arabic}

\section{Introduction}\label{sec:Introduction}

Reversible computation is an unconventional form of computing that uses reversible operations, 
that is, operations that can be easily and exactly reversed, or undone.  
Its study originates in the 1960's when it was observed
that maintaining physical reversibility avoids the dissipation of 
heat~\cite{Landauer}. Thus, the design of 
reversible logic gates and circuits can help to reduce the overall energy dissipation of 
computation and lead to low-power computing. Subsequently, motivation for 
studying reversibility has stemmed from a wide variety of applications
which  naturally embed reversible behaviour.  These include biological
processes where computation may be carried
out in forward or backward direction~\cite{ERK,LocalRev}, and
the field of system reliability
where reversibility can be used as a means of recovering from 
failures~\cite{TransactionsRCCS,LaneseLMSS13}.

One line of work in the study of reversible computation has been the investigation of its 
theoretical foundations. During the last few years a number of formal models have been
developed aiming to provide understanding of the basic principles of reversibility along with its
costs and limitations, and to explore
how it can be used to support the solution of complex problems. 
In the context of concurrent and distributed computation this study has led to the definition of 
different  forms of reversibility: While in the sequential setting reversibility 
is generally understood as the ability to execute past actions in the 
exact inverse order in which they have occurred, a process commonly referred
to as backtracking, in a concurrent scenario it can be argued that reversal
of actions can take place
in a more liberal fashion. The main alternatives proposed are those of
causal-order reversibility, a form of reversing where an action can be 
undone provided that all of its effects 
(if any) have been undone beforehand, and out-of-causal order reversibility, 
a form of reversing featured most notably in biochemical systems. 
These concepts have been studied 
within a variety of formalisms. 

To begin with, a large amount of work has focused on providing a formal understanding of 
reversibility within process calculi. The main challenge in these works has been
to maintain the information needed to reverse executed computation, e.g., to keep
track of the history of execution and the choices that have not been made. The first works 
handling reversibility in process calculi are
the Chemical Abstract Machine~\cite{abstract}, a calculus inspired by reactions between 
molecules whose operational semantics define both forward and reverse computations, and
RCCS~\cite{RCCS} an extension of  the Calculus of Communicating Systems (CCS)~\cite{CCS}
equipped with a reversible mechanism that uses memory stacks for concurrent 
threads, further developed in~\cite{TransactionsRCCS,BiologyCCSR}.  This mechanism was 
represented at an abstract level using categories with an application to Petri 
nets~\cite{DBLP:journals/entcs/DanosKS07}. Subsequently,  a general
method for reversing process calculi with CCSK being a special 
instance of the methodology was proposed in~\cite{Algebraic}. This proposal introduced the use
of communication keys to bind together 
communication actions as needed for isolating communicating partners during action reversal. 
Reversible versions of the $\pi$-calculus 
include  $\rho\pi$~\cite{LaneseMS10} and $R\pi$~\cite{DBLP:conf/lics/CristescuKV13}.
While all the above concentrate on the notion of causal reversibility, approaches considering
other forms of reversibility have also been proposed. In \cite{LocalRev} a new operator is 
introduced for modelling local reversibility,
a form of out-of-causal-order reversibility, whereas mechanisms for controlling 
reversibility have also been
considered in 
roll-$\pi$~\cite{LaneseMSS11}  and croll-$\pi$~\cite{LaneseLMSS13}.
The modelling of bonding within reversible processes and event structures 
was considered in~\cite{Bonding}, whereas a reversible computational calculus for modelling 
chemical systems composed of signals and gates was  proposed in~\cite{CardelliL11}.
The study of reversible process calculi has also triggered research on various
other models of concurrent computation such as reversible event 
structures~\cite{ConRev}.

In this work we focus on Petri nets (PNs)~\cite{PNs}, a graphical mathematical language 
that can be used for the specification and
analysis of discrete event systems. PNs are associated with 
a rich mathematical theory and a variety of tools, and they have been used extensively
for modelling and reasoning about a wide range of applications. 
The first study of reversible computation within PNs was proposed 
in~\cite{PetriNets,BoundedPNs}. In these works, the
authors investigate the effects of adding \emph{reversed} versions 
of selected transitions in a Petri net and they explore decidability
problems regarding reachability and coverability in the resulting Petri net. 
Through adding new transitions, this approach results in increasing the PN size and enabling 
reversal of not previously executed transitions (due to backward 
conflict in the structure of a Petri net). Addressing this issue,
our previous work of~\cite{RPNs}, introduces reversing Petri nets 
(RPNs), a variation of acyclic Petri nets where 
executed transitions can be reversed according to 
the  notions of \emph{backtracking}, 
\emph{causal reversibility} and \emph{out-of-causal-order reversibility}. 
A main feature of the formalism is that
during a transition firing, tokens can be bonded with each other. The creation of bonds
is considered to be the effect of a transition, whereas their destruction is the effect of
the transition's reversal. The proposal is motivated by applications from biochemistry,
but can be applied to a wide range of
problems featuring reversibility. 

It was shown that RPNs can be 
translated to bounded Coloured Petri Nets (CPNs), an extension of traditional
Petri Nets, where tokens carry data values, demonstrating that
the abstract model of RPNs, and thus the
principles of reversible computation, can be emulated in CPNs~\cite{RPNtoCPN}.
Furthermore, a mechanism for controlling reversibility in RPNs 
was proposed in~\cite{RC19}. This control is enforced with
the aid of conditions associated with transitions, whose satisfaction/violation acts as 
a guard for executing the transition
in the forward/backward direction, respectively. Recently, an alternative 
approach to that of RPNs~\cite{RPT} introduces reversibility in Petri
nets by unfolding the original Petri net into 
occurrence nets and coloured Petri nets. The authors encode causal memories 
while preserving the original computation by adding for each transition 
its reversible counterpart.

\paragraph{\bf Contribution.} In this work we present reversing Petri 
nets as a model that captures the main strategies for reversing computation,
i.e., \emph{backtracking}, \emph{causal reversibility}, and 
\emph{out-of-causal-order reversibility} within the Petri net framework.
The paper extends~\cite{RPNs} by introducing cycles into the framework 
and including the proofs of all results.
The extension of RPNs to the cyclic case turns out to be quite nontrivial 
since the presence of cycles
exposes the need to define causality of actions within a cyclic structure. 
Indeed there are different
ways of introducing reversible behaviour depending on how causality is defined. 
In our approach, we follow the 
notion of causality as defined by Carl Adam Petri  for one-safe nets that  
provides the notion of a run of a system where causal dependencies are 
reflected in terms of a partial order~\cite{PNs}. A causal
link is considered to exist between two transitions if one produces tokens 
that are used to fire the other. In this partial order, causal dependencies  
are explicitly defined as an unfolding of an occurrence net, which is 
an acyclic net that does not have backward conflicts.  We prove 
that the amount of flexibility allowed in causal reversibility indeed yields a 
causally-consistent semantics. We also demonstrate that out-of-causal-order  
reversibility is  able to create new states unreachable by forward-only execution which, nonetheless, only returns tokens to places that they have previously visited.
Additionally, we establish the relationship between the three forms of reversing and define
a transition relation that can capture each of the three strategies modulo the enabledness 
condition for each strategy. 
We demonstrate the framework with a model of the ERK pathway,
an example that inherently features out-of-causal-order reversibility.

\paragraph{\bf Paper organisation.}
In the next section we give an overview of the different types of 
reversibility and their characteristics and
we discuss the challenges of modelling reversibility in the context 
of Petri nets. In Section 3 we introduce the formalism of  
Reversing Petri nets including cycles. In Sections 4 and 5 we present semantics
and associated results for 
forward 
execution, backtracking, causal and out-of-causal-order 
reversibility, and we study
their relationship. In Section 6 we illustrate the framework with 
a model of the ERK pathway, and Section 7 concludes the paper.

\section{Forms of Reversibility and Petri Nets}\label{sec:Forms of Reversibility}

Reversing computational processes in concurrent and distributed systems has many
promising applications but also presents some  technical and conceptual challenges. 
In particular, a formal model for concurrent systems that embeds reversible computation needs to 
be able to compute without forgetting and 
to identify the legitimate backward moves at any point during computation.

The first challenge applies to both concurrent and sequential systems. Since
processes typically do not remember their past states, reversing their execution is not directly supported.
This challenge can be addressed with the use of memories. When building a reversible variant
of a formal language, its syntax can be extended to include appropriate representations for computation 
memories to allow processes to keep track of past execution.

The second challenge regards the strategy to be applied when going backwards.
Several approaches for performing and undoing steps 
have been explored in the literature over the last decade, which differ in the order
in which steps are taken backwards. The most prominent are \emph{backtracking}, 
\emph{causal reversibility}, and \emph{out-of-causal-order reversibility}.
{\em Backtracking} is well understood as the process of rewinding one's
computation trace, that is, computational steps 
are undone in the exact inverse order to the one in which they have occurred. 
This strategy ensures that at any state in a computation there is at most one predecessor 
state, yielding the property of \emph{backwards determinism}. 
In the context of reversible systems, this form of reversibility can be
thought of as overly restrictive since undoing moves only
in the order in which they were taken, induces fake causal
dependencies on backward sequences of actions: 
actions which could have been reversed in any order, e.g., 
actions belonging to independent threads,
are forced to be undone in the precise order in which they occurred.
To relax the rigidity of backtracking,  {\em causal reversibility} provides a more 
flexible form of reversing by allowing events to reverse in an 
arbitrary order, assuming that they respect the causal 
dependencies that hold between them.  Thus, in the context 
of causal reversibility, reversing does not have to follow the 
exact inverse order for events as long as caused actions, also
known as effects, are undone before 
the reversal of the actions that have caused them. A main
feature of causal reversibility is that
reversing an action returns a thread into a previously
executed state, thus, any continuation of 
the computation after the reversal would also be possible in 
a forward-only execution where the specific step was not taken in the
first place.

\begin{figure}
	\centering
	\subfigure{\includegraphics[width=5.5cm]{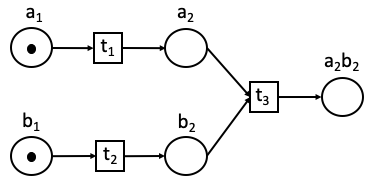}}
	\caption{ Causal Reversibility}
	\label{causal}
\end{figure}

For example consider the Petri net in Figure \ref{causal}.  We may observe
that transitions $t_1$ and $t_2$ are independent from each other as they may be taken in any
order, and they are both prerequisites for transition $t_3$. 
Backtracking the sequence of transitions  $\langle t_1, t_2,t_3\rangle$ would require that
the three transitions should be reversed in exactly the reverse order, i.e. 
$\langle t_3, t_2,t_1\rangle$. Instead, causal flexibility 
allows the inverse computation to rewind $t_3$ and then $t_1$ and $t_2$ in any order, but never
$t_1$ or $t_2$  before $t_3$.

Both backtracking and causal reversing are cause-respecting. There are, however, many important examples 
where concurrent systems violate causality since undoing things in an {\em out-of-causal} order is either inherent or
could be beneficial, e.g., biochemical reactions or mechanisms driving long-running transactions. 
This is due to the distinguishing characteristic of out-of-causal-order reversibility to
allow a system to discover states that are inaccessible in any forward-only execution. This
can be achieved since, reversing in out-of-causal order allows reversing an action before its effects are undone
and subsequently exploring new computations while the effects of the reversed action are still present.
As such, out-of-order reversibility can  create fresh alternatives of current states that 
were formerly inaccessible by any forward-only execution path.

\begin{figure}
	\centering
	\subfigure{\includegraphics[width=8cm]{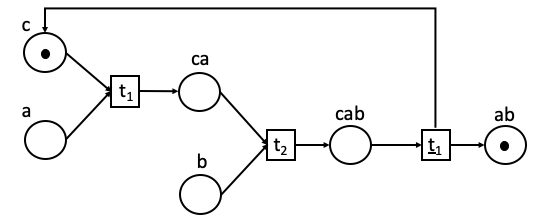}}
	\caption{ Catalysis in classic Petri nets}
	\label{catalyst2}
\end{figure}

Since out-of-order reversibility  contradicts program order, it comes with its 
own peculiarities that need to be taken into consideration while designing reversible systems. To appreciate
these peculiarities and obtain insights towards our approach on addressing reversibility within Petri nets, consider the process of catalysis from biochemistry, 
whereby a substance called \emph{catalyst} enables a chemical reaction between a set of other elements.  
Specifically consider a catalyst $c$ that helps the otherwise inactive molecules $a$  and $b$ to bond. This is achieved
as follows: catalyst $c$ initially bonds with $a$ which then enables the bonding 
between $a$ and $b$. Finally, the catalyst is no longer needed and its bond to
the other two molecules is released. A Petri net model of this process is illustrated in Figure~\ref{catalyst2}.
The Petri net executes transition $t_1$ via which the bond $ca$ is created, followed by
action $t_2$ to produce $cab$. Finally, action
$\underline{t_1}$ ``reverses'' the bond between $a$ and $c$, yielding $ab$ and releasing $c$. (The figure portrays the final state of the execution assuming that initially exactly one token existed in places $a$, $b$, and $c$.)

This example illustrates that Petri nets are not reversible by 
nature, in the sense that every transition cannot be executed
in both directions. Therefore an inverse action, (e.g., transition $\underline{t_1}$  for
undoing the effect of transition $t_1$) needs to be added as 
a supplementary forward transition for achieving the undoing of a previous action. 
This explicit approach of modelling reversibility can prove 
cumbersome in systems that express multiple reversible
patterns of execution, resulting in larger and more intricate systems. Furthermore, it
fails to capture reversibility as a mode of computation.
The intention of our work is to study an approach for modelling 
reversible computation that does not require the addition of new, 
reversed transitions but
instead offers as a basic building block transitions that can 
be taken in both the forward as well as the backward direction, and, 
thereby, explore the theory
of reversible computation within Petri nets.

However, when attempting to model the catalysis example
while executing transitions in both
the forward and the backward directions, we may observe a 
number of obstacles. At an abstract level, the behaviour of 
the system should exhibit a sequence of three transitions: 
execution of $t_1$ and $t_2$, followed by the reversal of transition $t_1$. 
The reversal of transition $t_1$ should implement the release of $c$ from the bond $cab$ and make
it available for further instantiations of transitions, 
if needed, while the bond $ab$ should remain in place.
This implies that a reversing Petri net model should provide resources
$a$, $b$ and $c$ as well as $ca$, $cab$ and $ab$ and 
implement the reversal of action $t_1$ as the
transformation of resource $cab$ into $c$ and $ab$. Note that resource $ab$ is inaccessible during the forward execution 
of transitions $t_1$ and $t_2$ and only materialises after the 
reversal of transition $t_1$,
i.e., only once the bond between $a$ and $c$ is broken. Given 
the static nature of a Petri net, this suggests that
resources such as $ab$ should be  represented at the token
level (as opposed to the place level).    
As a result, the concept of token individuality is of particular relevance to
reversible computation in Petri nets while other constructs/functions 
at token level are needed to capture the effect and reversal
of a transition.

Indeed, reversing a transition in an out-of-causal order may imply 
that while some of the effects of the transition
can be reversed (e.g., the release of the catalyst back to the 
initial state), others must be retained due to computation that 
succeeded the forward execution of the next transition (e.g., 
token $a$ cannot be released during the reversal of $t_1$ since
it has bonded with $b$ in transition $t_2$). This latter point 
is especially challenging since it requires to specify a model 
in a precise manner so as to identify which effects are allowed 
to be ``undone'' when reversing a transition. 
Thus, as highlighted by the catalysis example, reversing transitions 
in a Petri net model requires close monitoring of
token manipulation within a net and clear enunciation of the effects of a transition.

\remove{
	\begin{figure}
		\centering
		\subfigure{\includegraphics[width=8cm]{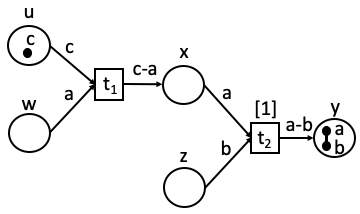}}
		\caption{ Catalysis in reversing Petri nets}
		\label{catalyst}
	\end{figure}
}

As already mentioned, the concept of token individuality can 
prove useful to handle these challenges.
This concept has also been handled in various works, 
e.g.,~\cite{Causal,Individual,Glabbeek},
where each token is associated with information regarding its 
causal path, i.e., the places and transitions
it has traversed before reaching its current state.  
In our approach, we also implement  the notion of token individuality 
where instead of maintaining extensive histories for recording
the precise evolution of each token through transitions and places, 
we  employ a novel approach inspired
by out-of-causal reversibility in biochemistry as well as approaches 
from related literature~\cite{Bonding}. The resulting 
framework is light in
the sense that no memory needs to be stored per token to retrieve its causal path
while enabling reversible semantics for the three main types of reversibility. 
Specifically, 
we introduce two notions that intuitively capture tokens and their history: 
the notion of a \emph{base} and a new type of tokens called \emph{bonds}. 
A base is a persistent type of token which cannot be
consumed and therefore preserves its individuality through various transitions. 
For a transition to fire, the incoming arcs identify the required tokens/bonds 
and the outgoing arcs may create new bonds or transfer already existing 
tokens/bonds along the places of a PN.  
Therefore, the effect of a transition is the creation of new \emph{bonds} 
between the tokens it takes as input and the reversal of such a transition 
involves undoing the respective bonds. 
In other words, a token can be a base or a  coalition of bases 
connected via bonds into a structure. 

\begin{figure}
	\centering
	\subfigure{\includegraphics[width=6cm]{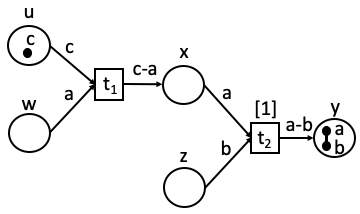}}
	\caption{ Catalysis in reversing Petri nets}
	\label{catalyst}
\end{figure}

Based on these ideas, we may describe the catalysis example in 
our proposed framework as shown in Figure~\ref{catalyst}. 
In this setting $a$, and $c$ are bases which  are connected via a 
bond into place $x$ during transition $t_1$, while transition 
$t_2$ brings into place a new bond between $a$ and $b$.  In Figure~\ref{catalyst} 
we see the state that arises after the execution of $t_1$ and $t_2$ and the reversal 
of transition $t_1$. In this state, base $c$ has returned to its initial 
place $u$ whereas bond $a-b$ has remained in place $y$. A thorough 
explanation of the notation is given in the next section.

Finally, in order to identify at each point in time the
history of execution, thus to discern the transitions that can be reversed
given the presence of backward nondeterminism of Petri nets, we associate 
transitions with a history storing keys in increasing order each time an 
instance of the transition is executed. This allows to backtrack computation 
as well as to extract the causes of bonds as
needed in causal and out-of-causal-order reversibility.

\section{Reversing Petri Nets}\label{sec: Reversible Petri Nets}

We define reversing Petri nets as follows:

\begin{definition}{\rm
		A \emph{\PN}(RPN) is a tuple $(A, P,B,T,F)$ where:
		\begin{enumerate}
			\item $A$ is a finite set of \emph{bases} or \emph{tokens} ranged over by $a$, $b$, $\ldots$  $\overline{A} = \{\overline{a}\mid a\in A\}$
			contains a \emph{negative instance} for each token and we write ${\cal{A}}=A \cup \overline{A}$.
			\item $P$ is a finite set of \emph{places}.
			\item $B\subseteq A\times A$ is a set of undirected \emph{bonds} ranged over by $\beta$, $\gamma$, $\ldots$
			We use the notation $a \bond b$ for a bond $(a,b)\in B$. $\overline{B} = \{\overline{\beta}\mid \beta\in B\}$
			contains a \emph{negative instance} for each bond and we write ${\cal{B}}=B \cup \overline{B}$.
			\item $T$ is a finite set of \emph{transitions}.
			\item $F : (P\times T  \cup T \times P)\rightarrow 2^{{\cal{A}}\cup {\cal{B}}}$ defines a set of directed \emph{arcs} each associated with a subset of ${\cal{A}}\cup {\cal{B}}$.
		\end{enumerate}
}\end{definition}

A \PN is built on the basis of a set of \emph{bases} or \emph{tokens}. We 
consider each token to have a unique name. In this way, tokens may be distinguished from
each other, their persistence can be guaranteed and their history inferred from the structure
of a Petri net (as implemented by function $F$, discussed below). Tokens correspond to
the basic entities that occur in a system. 
They may occur as stand-alone elements but they may also merge together to form \emph{bonds}. \emph{Places}
and \emph{transitions} have the standard meaning. 

Directed arcs connect places to transitions and vice
versa and are labelled by a subset of ${\cal{A}}\cup {\cal{B}}$ where  $\overline{A} = \{\overline{a}\mid a\in A\}$
is the set of \emph{negative} tokens expressing token absence, and  $\overline{B} = \{\overline{\beta}\mid \beta\in B\}$ is the set of \emph{negative} bonds expressing bond absence.
For a label $\ell= F(x,t)$ or $\ell = F(t,x)$, we assume that each token $a$ 
can appear in $\ell$ at most once, either as $a$ or as $\overline{a}$,
and that if a bond $(a,b)\in\ell$ then $a,b\in\ell$. Furthermore, for $\ell = F(t,x)$,
it must be that $\ell\cap (\overline{A}\cup \overline{B}) = \emptyset$, that is, negative tokens/bonds may only occur
on arcs incoming to a transition. 
Intuitively, these labels express the requirements for a transition
to fire when placed on arcs incoming the transition, and the effects of the transition when placed on the
outgoing arcs. Thus, if $a\in F(x,t)$ this implies that token $a$ is required for the transition $t$ to
fire, and similarly for a bond $\beta\in F(x,t)$. On the other hand, $\overline{a}\in F(x,t)$ expresses that
token $a$ should not be present in the incoming place $x$ of $t$ for the transition to fire and similarly for a bond $\beta$, 
$\overline\beta \in F(x,t)$. Note
that negative tokens/bonds are close in spirit to the inhibitor arcs of extended Petri nets. Finally, note that 
$F(x,t)= \emptyset$ implies that there is no arc from place $x$ to transition $t$ and similarly for
$F(t,x) = \emptyset$.

We introduce the following notations. We write 
$\circ t =   \{x\in P\mid  F(x,t)\neq \emptyset\}$ and  
$ t\circ = \{x\in P\mid F(t,x)\neq \emptyset\}$
for the incoming and outgoing places of transition
$t$, respectively. Furthermore, we write
$\guard{t}  =   \bigcup_{x\in P} F(x,t)$ for the union of all labels on the incoming arcs of  transition $t$, and
$\effects{t}  =   \bigcup_{x\in P} F(t,x)$ for the union of all labels on the outgoing arcs of transition $t$.
\begin{definition}\label{well-formed}{\rm 
		A \PN $(A,P,B,T,F)$ is \emph{well-formed} if it satisfies the following conditions for all $t\in T$:
		\begin{enumerate}
			\item $A\cap \guard{t} = A\cap \effects{t}$,
			\item If $ a \bond b \in \guard{t}$ then $ a \bond b \in \effects{t}$,
			\item $ F(t,x)\cap F(t,y)=\emptyset$ for all $x,y\in P$, $x\neq y $. 
		\end{enumerate}
}\end{definition}

According to the above we have that: (1) transitions do not
erase tokens, (2) transitions do not destroy bonds, that is, if a bond $a\bond b$ exists in an input place of a transition, then it is
maintained in some output place, and 
(3) tokens/bonds cannot be cloned into more than one outgoing place.

As with standard Petri nets, we employ the notion of a \emph{marking}. A marking is a distribution
of tokens and bonds across places,  $M: P\rightarrow 2^{A\cup B}$ where $a \bond b \in M(x)$, for some $x\in P$, implies
$a,b\in M(x)$. 
In addition, we employ the notion of a \emph{history}, which assigns a memory to each transition of a reversing Petri net as
$H : T\rightarrow 2^\mathbb{N}$. Intuitively, a history of $H(t) = \emptyset$ for some $t \in T$ captures that the transition
has not taken place, and a history of 
$H(t) = \{k_1,\ldots,k_n\}$ captures that the transition was executed
and not reversed $n$ times where $k_i$, $1\leq i \leq n$, indicates the order of execution of the $i^{th}$ instance
amongst non-reversed actions. Note that this machinery, extending the exposition of~\cite{RPNs},
is needed to accommodate the presence of cycles, which yield the
possibility of repeatedly executing the same transitions.
$H_0$ denotes the initial history where $H_0(t) = \emptyset$ for all $t\in T$. 
A pair of a marking and a history describes a \emph{state} of a PN based on which execution is determined. 
We use the notation $\state{M}{H}$ to denote states. 

In a graphical representation, tokens are indicated by $\bullet$, places by circles, transitions by boxes, and bonds by lines 
between tokens. Furthermore,
histories are presented over the respective transitions as the list  $[k_1,...,k_n]$ when
$H(t) = \{k_1,\ldots,k_n\}$, $n>0$, and omitted when $H(t)=\emptyset$. 

As the last piece of our machinery, we define a notion that identifies connected
components of tokens and their associated bonds within a place. 
Note that more than one connected component may arise
in a place due to the fact that various unconnected tokens may be moved to a place 
simultaneously by a transition, while the reversal of transitions, which results
in the destruction of bonds, may break down a connected component into various
subcomponents. We define $\connected(a,C)$, where $a$ is a base and $C\subseteq A\cup B$ to be the tokens connected
to $a$ via sequences of bonds as well as the bonds creating these connections according to 
set $C$.  
\[\connected(a,C)=(\{a\} \cap C)\cup\{\beta, b, c \mid \exists  w \mbox{ s.t. }  \paths(a,w,C), \beta\in w \mbox{, and } \beta=(b,c) \}\]
where $\paths(a,w,C)$ if $w=\langle\beta_1,\ldots,\beta_n\rangle$, and for all $1\leq i \leq n$, $\beta_i=(a_{i-1},a_i)\in C\cap B$, $a_i\in C\cap A$, and $a_0 = a$.

Returning to the example of Figure~\ref{catalyst}, we may see
a reversing net with three tokens $a$, $b$, and $c$, transition $t_1$, which bonds tokens $a$ and $c$ within place $x$,
and  transition $t_2$, which bonds the $a$ of bond $c \bond a$ with token $b$ into place $y$.
{Note that to avoid overloading figures, we omit writing the bases of
	bonds on the arcs of RPNs, so, e.g., on the arc between $t_1$ and $x$, we write $a \bond b$ as opposed to $\{a \bond b, a, b\}$.}
(The marking depicted in the figure is the one arising after the execution of transitions $t_1$ and $t_2$ and subsequently 
the reversal of transition $t_1$ by the semantic relations to be defined in the next section.) 

We may now define the various types of execution for  reversing Petri nets. In what follows we
restrict our attention to  well-formed RPNs $(A,P,B,T,F)$ with initial
marking $M_0$ such that for all $a\in A$, $|\{ x \mid a\in M_0(x)\} | = 1$.

\section{Forward Execution}
In this section we consider the standard, forward execution of RPNs.
\begin{definition}\label{forward}{\rm
		Consider a reversing Petri net $(A, P,B,T,F)$, a transition $t\in T$, 
		and a state $\state{M}{H}$. We say that
		$t$ is \emph{forward-enabled} in $\state{M}{H}$  if the following hold:
		\begin{enumerate}
			\item  if $a\in F(x,t)$,  for some $x\in\circ t$, then $a\in M(x)$, and if   
			$\overline{a}\in F(x,t)$
			for some $x\in\circ t$, then $a\not\in M(x)$, 
			\item  if $\beta\in F(x,t)$,  for some $x\in\circ t$, then 
			$\beta\in M(x)$, and if $\overline{\beta}\in F(x,t)$
			for some $x\in\circ t$, then $\beta\not\in M(x)$, 
			\item if $a\in F(t,y_1)$,  $b\in F(t,y_2)$, $y_1 \neq y_2$, then $b \not\in \connected(a,M(x))$ for all
			$x \in \circ t$, and  
			\item if $\beta\in F(t,x)$ for some $x\in t\circ$ and $\beta\in M(y)$ for some $y\in \circ t$, then $\beta\in F(y,t)$. 
		\end{enumerate}
}\end{definition}

Thus, $t$ is enabled in state $\state{M}{H}$ if (1), (2), all 
tokens and bonds required for the transition to take place
are available in the incoming places of $t$ and  none of the 
tokens/bonds whose absence is
required exists in an incoming place of the transition, (3) if 
a transition forks into outgoing places $y_1$ and
$y_2$ then the tokens transferred to these places are not connected to each other in the
incoming places of the transition,  and (4) if a pre-existing bond 
appears in an outgoing arc of a transition, then it is also a precondition 
of the transition to fire. 
Contrariwise, if the bond appears in an outgoing arc of a 
transition ($\beta\in F(t,x)$ for some $x\in t\circ$)
but is not a requirement for the transition to fire ($\beta\not\in F(y,t)$ for all $y\in \circ t$),
then  the bond should not be present in an incoming place of the transition ($\beta\not\in M(y)$ for all $y\in \circ t$).

We observe that the new bonds created by a transition are
exactly those that occur in the outgoing edges of a transition but not in the incoming edges. Thus, we define the effect of a transition as 
\[\effect{t} = \effects{t} - \guard{t}\]
This will subsequently enable the enunciation of transition reversal by the destruction of exactly
the bonds in $\effect{t}$.

\begin{definition}{\rm \label{forw}
		Given a reversing Petri net $(A, P,B,T,F)$, a state $\langle M, H\rangle$, and a transition $t$ enabled in 
		$\state{M}{H}$, we write $\state{M}{H}
		\trans{t} \state{M'}{H'}$
		where:
		\[
		\begin{array}{rcl}
		M'(x) & = & \left\{
		\begin{array}{ll}
		M(x)-\bigcup_{a\in F(x,t)}\connected(a,M(x))  & \textrm{if } x\in \circ{t} \\
		M(x)\cup F(t,x)\cup \bigcup_{ a\in F(t,x)\cap F(y,t)}\connected(a,M(y)) 
		& \textrm{if }  x\in t\circ\\
		M(x), &\textrm{otherwise}
		\end{array}
		\right.
		\end{array}
		\]
		and
		\[
		\begin{array}{rcl}
		H'(t') & = & \left\{
		\begin{array}{lll}
		H(t')\cup \{ \max( \{0\} \cup\{k \mid k\in H(t''), t''\in T\}) +1\},\hspace{0.1in}
		\textrm{if } t' = t\\
		H(t'),\hspace{3in}\textrm{ otherwise}
		\end{array}
		\right.
		\end{array}
		\]
}\end{definition} 

\begin{figure}[t]
	\centering
	\subfigure{\includegraphics[width=5cm]{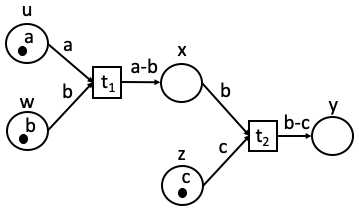}}
	\subfigure{\includegraphics[width=.6cm]{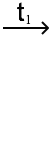}}
	\subfigure{\includegraphics[width=5cm]{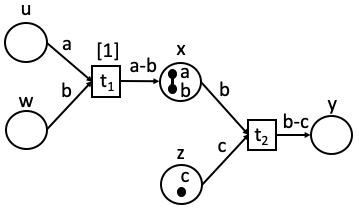}}\\
	\subfigure{\includegraphics[width=.6cm]{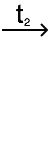}}
	\subfigure{\includegraphics[width=5cm]{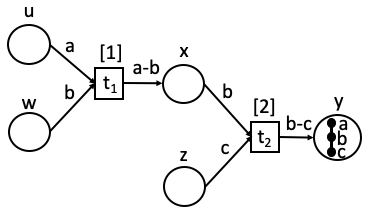}}
	\caption{Forward execution}
	\label{f-example}
\end{figure}

Thus, when a transition $t$ is executed in {the} forward direction, all tokens and bonds
occurring in its incoming arcs are relocated from the input 
places to the output places along with their connected 
components. An example of forward transitions can be seen in 
Figure ~\ref{f-example} where transitions $t_1$ and $t_2$ take 
place with the histories of the two transitions 
becoming $[1]$ and $[2]$, respectively. 

We may prove the following result, which verifies that bases are preserved during forward execution in the sense that transitions neither erase nor clone them. As far as bonds are concerned, the proposition states that forward execution may
create but not destroy bonds.
\begin{proposition} [Token and bond preservation]  \label{prop1}    {\rm Consider a reversing Petri net $(A, P,B,T, F)$, a state $\langle M, H\rangle$ such that
		for all $a\in A$,  $|\{x\in P\mid a\in M(x)\}| = 1$, and a transition 
		$\state{M}{H} \trans{t} \state{M'}{H'}$. Then:
		\begin{enumerate}
			\item for all $a\in A$,  $|\{x\in P \mid a\in M'(x)\}|=1$,
			and 
			\item for all $\beta\in B$, $|\{x\in P \mid \beta\in M(x)\}| \leq |\{x\in P \mid \beta\in M'(x)\}|\leq 1$.
		\end{enumerate}
}\end{proposition}

\paragraph{Proof:}
The proof of the result follows the definition of forward execution and relies on the well-formedness of
RPNs. Consider a reversing Petri net $(A, P,B,T,F)$, a state $\langle M, H\rangle$  such that $|\{x\in P\mid a\in M(x)\}| = 1$ for
all $a\in A$, and suppose 
$\state{M}{H} \trans{t} \state{M'}{H'}$.  

For the proof of clause (1) let $a\in A$. Two cases exist:
\begin{enumerate}
	\item $a\in \connected(b,M(x))$ for some $b\in F(x,t)$. Note that $x$ is unique by the
	assumption that $|\{x\in P\mid a\in M(x)\}| = 1$. Furthermore, according to Definition~\ref{forw}, 
	we have that $M'(x) = M(x) - \{\connected(b,M(x)) \mid b\in F(x,t)\}$, which implies that $a\not\in M'(x)$.
	On the other hand,  by  Definition~\ref{well-formed}(1),
	$b\in \effects{t}$. Thus, there exists $y\in t\circ$, such that $b\in F(t,y)$. Note that this $y$
	is unique by Definition~\ref{well-formed}(3).  As a result, by Definition~\ref{forw}, 
	$M'(y) = M(y)\cup F(t,y)\cup \{\connected(b,M(x)) \mid b\in F(t,y), x\in\circ{t}\}$. Since
	$b\in F(x,t)\cap F(t,y)$, $a\in \connected(b,M(x))$, this implies that $a\in M'(y)$. 
	
	Now suppose that $a\in \connected(c,M(x))$ for some  $c\neq b$, $c\in F(t,y')$. Then, by Definition~\ref{forward}(3), 
	it must be that $y = y'$. As a result, we have that $\{z\in P\mid a\in M'(z)\} = \{y\}$ and 
	the result follows.
	\item $a\not\in \connected(b,M(x))$ for all $b\in F(x,t)$, $x\in P$. This implies that 
	$\{x\in P\mid a\in M'(x)\} = \{x\in P\mid a\in M(x)\}$ and the result follows.
\end{enumerate}

To prove clause (2) of the proposition, consider a bond
$\beta\in B$, $\beta=(a,b)$. We observe that, since $|\{x\in P\mid a\in M(x)\}| = 1$ for
all $a\in A$, $|\{x\in P\mid \beta\in M(x)\}| \leq 1$. The proof follows by case analysis as
follows:
\begin{enumerate}
	\item Suppose $|\{x\in P\mid \beta\in M(x)\}| =0$. Two cases exist:
	\begin{itemize}
		\item Suppose $\beta\not\in F(t,x)$ for all $x\in P$. Then, by 
		Definition~\ref{forw}, $\beta\not\in M'(x)$ 
		for all $x\in P$. Consequently, $|\{x\in P\mid \beta\in M'(x)\}| =0$ and the result follows.
		\item Suppose $\beta\in F(t,x)$ for some $x\in P$. Then, by  Definition~\ref{well-formed}(3),
		$x$ is unique, and by Definition~\ref{forw}, $\beta\in M'(x)$.
		Consequently, $|\{x\in P\mid \beta\in M'(x)\}| =1$ and the result follows.
	\end{itemize}
	\item Suppose $|\{x\in P\mid \beta\in M(x)\}| =1$. Two cases exist:
	\begin{itemize}
		\item $\beta\not\in \connected(c,M(x))$ for all $c\in F(x,t)$. 
		This implies that $\{x\in P\mid \beta\in M'(x)\} = 
		\{x\in P\mid \beta\in M(x)\}$ and the result follows.
		\item $\beta\in \connected(c,M(x))$ for some $c\in F(x,t)$. Then, according to Definition~\ref{forw}, 
		we have that $M'(x) = M(x) - \{\connected(c,M(x)) \mid c\in F(x,t)\}$, which implies that $\beta\not\in M'(x)$.
		On the other hand, by the definition of well-formedness, Definition~\ref{well-formed}(1),
		$c\in \effects{t}$. Thus, there exists $y\in t\circ$, such that $c\in F(t,y)$. Note that this $y$
		is unique by Definition~\ref{well-formed}(3).  As a result, by Definition~\ref{forw}, 
		$M'(y) = M(y)\cup F(t,y)\cup \{\connected(c,M(x)) \mid c\in F(t,y), x\in\circ{t}\}$. Since
		$c\in F(x,t)\cap F(t,y)$, $\beta\in \connected(c,M(x))$, this implies that $\beta\in M'(y)$. 
		
		Now suppose that $\beta\in \connected(d,M(x))$ for some $d\neq c$, $c\in F(d,y')$. Then, by Definition~\ref{forward}, 
		and since $\connected(c,M(x)) = \connected(d,M(x))$,
		it must be that $y = y'$. As a result, we have that $\{z\in P\mid \beta\in M'(z)\} =  \{y\}$ and 
		the result follows.
	\end{itemize}
	\proofend
\end{enumerate}

\section{Reverse Execution}
\subsection{Backtracking}

Let us now proceed to the simplest form of reversibility, namely, backtracking. 
We define a transition to
be \emph{bt}-enabled (backtracking-enabled) if it was the last executed transition:

\begin{definition}\label{bt-enabled}{\rm
		Consider a state $\state{M}{H}$ and a transition $t\in T$. We say that $t$ is \emph{$bt$-enabled} in
		$\state{M}{H}$ if
		$k\in H(t)$ with $k\geq k'$ for all $k' \in H(t')$, $t'\in T$.
}\end{definition}

Thus, a transition $t$ is $bt$-enabled if its history contains the highest value among all transitions. 
The effect of backtracking a transition in a reversing Petri net is as follows:

\begin{definition}\label{br-def}{\rm
		Given a reversing Petri net $(A, P,B,T,F)$, a state $\langle M, H\rangle$, and a transition $t$ that is $bt$-enabled in $\state{M}{H}$, we write $ \state{M}{H}
		\btrans{t} \state{M'}{H'}$
		where:
		\[
		\begin{array}{rcl}
		M'(x) & = & \left\{
		\begin{array}{ll}
		M(x)\cup\bigcup_{ y \in t\circ, a\in F(x,t)\cap F(t,y)}\connected(a,M(y)-\effect{t}),  & \textrm{if } x\in \circ{t} \\
		M(x)- \bigcup_{a\in F(t,x)}\connected(a,M(x)) , & \textrm{if }  x\in t\circ\\
		M(x) &\textrm{otherwise}
		\end{array}
		\right.
		\end{array}
		\]
		and 
		\[
		\begin{array}{rcl}
		H'(t') & = & \left\{
		\begin{array}{lll}
		H(t')- \{ k \},\hspace{0.8in} & \textrm{if } t' = t, k = \max(H(t))\\
		H(t')  & \textrm{otherwise}
		\end{array}
		\right.
		\end{array}
		\]
		
}\end{definition}

\begin{figure}[t]
	\centering
	\subfigure{\includegraphics[width=5cm]{backtracking.png}}
	\subfigure{\includegraphics[width=.62cm]{arrow1.png}}
	\subfigure{\includegraphics[width=.6cm]{arrow2.png}}
	\subfigure{\includegraphics[width=5cm]{backtracking2.png}}\\
	\subfigure{\includegraphics[width=.6cm]{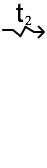}}
	\subfigure{\includegraphics[width=5cm]{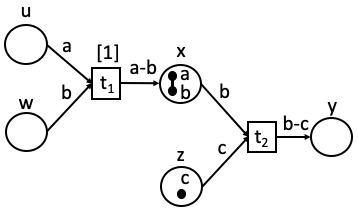}}
	\subfigure{\includegraphics[width=.6cm]{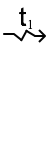}}
	\subfigure{\includegraphics[width=5cm]{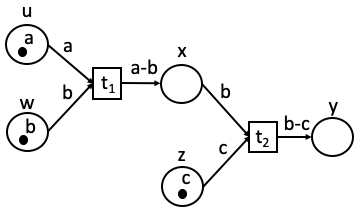}}
	\caption{Backtracking execution}
	\label{b-example}
\end{figure}

When a transition $t$ is reversed in a backtracking fashion  all tokens and bonds in the
postcondition of the transition, as well as their connected components, 
are transferred to the incoming places of the transition and any newly-created bonds are broken. 
Furthermore, the largest key in the history of the transition is removed.

\begin{figure}[t]
	\centering
	\subfigure{\includegraphics[width=5cm]{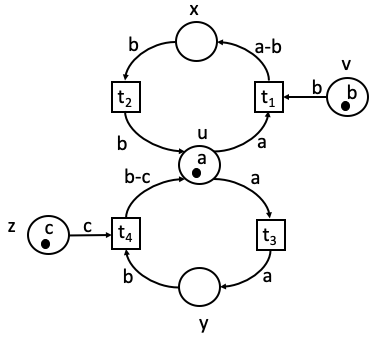}}
	\subfigure{\includegraphics[width=.5cm]{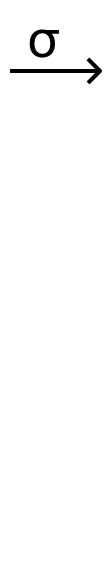}}
	\subfigure{\includegraphics[width=5cm]{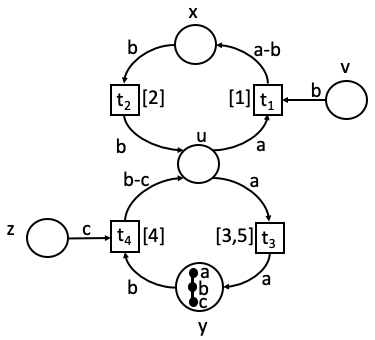}}\\
	\subfigure{\includegraphics[width=.5cm]{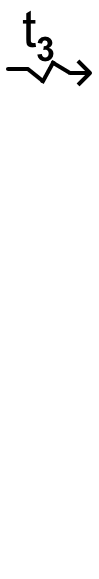}}
	\subfigure{\includegraphics[width=5cm]{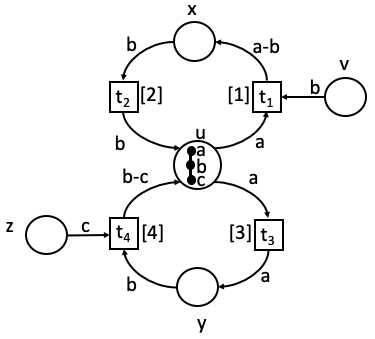}}
	\caption{Backtracking execution where $\sigma =\langle t_1,t_2,t_3,t_4,t_3\rangle$}
	\label{b-cycles}
\end{figure}

An example of backtracking extending the example of Figure~\ref{f-example} can be seen in  Figure~\ref{b-example} where we observe transitions $t_2$ and $t_1$ being reversed  with the histories of the two transitions being eliminated.
A further example can be seen in Figure~\ref{b-cycles} where after the execution of transition sequence $\langle t_1, t_2,t_3,t_4,t_3\rangle$, only transition $t_3$ is $bt$-enabled since it was the last transition to be executed. During its reversal, the component $a \bond b\bond c$ is returned to  place $u$. Furthermore, the largest key of the history of $t_3$ becomes empty.

We may prove the following result, which verifies that bases are preserved during backtracking execution in the sense that
there exists exactly one instance of each base and backtracking transitions neither erase nor clone them. 
As far as bonds are concerned, the proposition states that at any time there may exist at most one 
instance of a bond and that backtracking transitions may only destroy bonds.
\begin{proposition}  [Token preservation and bond destruction] \label{prop2}
	{\rm Consider a reversing Petri net $(A, P,B,T,F)$, a state $\langle M, H\rangle$
		such that for all $a\in A$, $|\{x\in P \mid a\in M(x)\}| = 1$, and a transition 
		$\state{M}{H} \btrans{t} \state{M'}{H'}$. Then: 
		\begin{enumerate} 
			\item for all $a\in A$,  
			$|\{x\in P \mid a\in M'(x)\}| = 1$, and 
			\item  for all $\beta\in B$, $1 \geq |\{x \in P \mid \beta\in M(x)\}| \geq |\{x \in P \mid \beta\in M'(x)\}|$.
		\end{enumerate}
}\end{proposition}
\paragraph{Proof:}
The proof of the result follows the definition of backward 
execution and relies on the well-formedness of
reversing Petri nets. 
Consider \RPN $(A, P,B,T,F)$, a state $\langle M, H\rangle$ such that $|\{x\in P \mid a\in M(x)\}| =1$
for all $a\in A$, and suppose  $\state{M}{H} \btrans{t} \state{M'}{H'}$.  

We begin with the proof of clause (1) and let $a\in A$. Two cases exist:
\begin{enumerate}
	
	\item $a\in \connected(b,M(x))$ for some $b\in F(t,x)$. Note that by the assumption 
	of $|\{x\in P \mid a\in M(x)\}| =1$, $x$ must be unique. 
	Let us choose $b$ such that, additionally, $a\in \connected(b,M(x) - \effect{t})$. Note
	that such a $b$ must exist, otherwise 
	the forward execution of $t$ would not have transferred $a$ along with $b$ to place $x$.
	
	According to Definition~\ref{br-def}, 
	we have that $M'(x) = M(x) - \{\connected(b,M(x)) \mid b\in F(t,x)\}$, 
	which implies that $a\not\in M'(x)$.
	On the other hand, note that by the definition of well-formedness, Definition~\ref{well-formed}(1),
	$b\in \guard{t}$. Thus, there exists $y\in \circ t$, such that $b\in F(y,t)$. Note that this $y$
	is unique. If not, then there exist
	$y$ and $y'$ such that $y\neq y'$ with $b\in F(y,t)$ and $b\in F(y',t)$.  By the assumption, however,
	that there exists at most one token of each base, and Proposition~\ref{prop1}, $t$ would never be enabled, 
	which leads to a contradiction.   As a result, by Definition~\ref{br-def}, 
	$M'(y) = M(y)\cup\{\connected(b,M(x)-\effect{t}) \mid b\in F(y,t)\cap F(t,x)\}$. Since
	$b\in F(y,t)\cap F(t,x)$, $a\in \connected(b,M(x)-\effect{t})$, this implies that $a\in M'(y)$. 
	
	Now suppose that $a\in \connected(c,M(x)-\effect{t})$, $c\neq b$, and $c\in F(y',t)$. Since
	$a\in \connected(b,M(x) - \effect{t})$, it must be that
	$\connected(b,M(x) - \effect{t})=\connected(c,M(x)-\effect{t})$. Since $b$ and $c$ are
	connected to each other but the connection was not created by transition $t$ (the connection is
	present in $M(x)-\effect{t}$), it must be that the connection was already present before
	the forward execution of $t$ and, by token uniqueness, we conclude that  $y=y'$.
	\item $a\not\in \connected(b,M(x))$ for all $b\in F(t,x)$, $x\in P$. This implies that 
	$\{x\in P\mid a\in M'(x)\} = \{x\in P\mid a\in M(x)\}$ and the result follows.
\end{enumerate}

Let us now prove clause (2) of the proposition. Consider a bond
$\beta\in B$, $\beta=(a,b)$. We observe that, since $|\{x\in P\mid a\in M(x)\}| = 1$ for
all $a\in A$, $|\{x\in P\mid \beta\in M(x)\}| \leq 1$. The proof follows by case analysis as
follows:
\begin{enumerate}
	\item $\beta\in \connected(c,M(x))$ for some $c\in F(t,x)$, $x\in P$. By the assumption 
	of $|\{x\in P \mid \beta\in M(x)\}| =1$, $x$ must be unique. 
	Then, according to Definition~\ref{br-def}, 
	we have that $M'(x) = M(x) - \{\connected(c,M(x)) \mid c\in F(x,t)\}$, 
	which implies that $\beta\not\in M'(x)$.
	Two cases exist: 
	\begin{itemize}
		\item If $\beta\in \effect{t}$, then $\beta\not\in M'(y)$ for all places $y\in P$.
		\item If $\beta\not\in \effect{t}$ then let us choose $c$ such that 
		$\beta\in \connected(c,M(x) - \effect{t})$. Note
		that such a $c$ must exist, otherwise 
		the forward execution of $t$ would not have connected $\beta$ with $c$. 
		By the definition of well-formedness, Definition~\ref{well-formed}(1),
		$c\in \guard{t}$. Thus, there exists $y\in \circ t$, such that $c\in F(y,t)$. Note that this $y$
		is unique (if not, $t$ would not have been enabled).  As a result, by Definition~\ref{br-def}, 
		$\beta \in M'(y)$.
		
		Now suppose that $\beta\in \connected(d,M(x)-\effect{t})$, $d\neq c$, and $d\in M'(y')$. Since
		$\beta\in \connected(c,M(x) - \effect{t})$, it must be that
		$\connected(c,M(x) - \effect{t})=\connected(d,M(x)-\effect{t})$.  Since $c$ and $d$ are
		connected to each other but the connection was not created by transition $t$ (the connection is
		present in $M(x)-\effect{t}$), it must be that the connection was already present before
		the forward execution of $t$ and, by token uniqueness, we conclude that  $y=y'$.
		This implies that $\{z\in P\mid \beta\in M'(z)\} = \{y\}$. 
	\end{itemize}
	The above imply that $\{z\in P\mid \beta\in M(z)\} = \{x\}$ and  $\{z\in P\mid \beta\in M'(z)\} \subseteq \{y\}$ and 
	the result follows.
	\item $\beta\not\in \connected(c,M(x))$ for all $c\in F(t,x)$, $x\in P$. This implies that 
	$\{x\in P\mid \beta\in M'(x)\} = \{x\in P\mid \beta\in M(x)\}$ and the result follows.
	\proofend
\end{enumerate}

Let us now consider the combination of forward and backward moves in executions. 
We write $\fbtrans{}$ for $\trans{}\cup\btrans{}$.
The following result establishes that in an execution beginning in the initial state of a reversing Petri net, bases are 
preserved, bonds can have at most one instance at any time and a new occurrence of  a bond may be created 
during a forward  transition that features the bond as its effect whereas a bond can be destroyed 
during the backtracking of a transition that features the bond as its effect. This last point clarifies
that the effect of a transition characterises the bonds that are newly-created during the transition's forward execution
and the ones that are destroyed during its reversal.

\begin{proposition}\label{Prop}{\rm Given a reversing Petri net $(A, P,B,T,F)$, an initial state 
		$\langle M_0, H_0\rangle$ and an execution
		$\state{M_0}{H_0} \fbtrans{t_1}\state{M_1}{H_1} \fbtrans{t_2}\ldots \fbtrans{t_n}\state{M_n}{H_n}$, the following hold:
		\begin{enumerate}
			\item For all $a\in A$  and $i$, $0\leq i \leq n$,   $|\{x\in P \mid a\in M_i (x)\}| = 1$.
			\item For all $\beta \in B$ and $i$, $0\leq i \leq n$, 
			\begin{enumerate}
				\item $0 \leq |\{x \in P \mid \beta\in M_i(x)\}| \leq 1$,
				\item if $t_i$ is executed in the forward direction and $\beta\in \effect{t_i}$, then 
				$\beta\in M_{i}(x)$ for some $x\in P$ where $\beta\in F(t_i,x)$, and  $\beta\not\in M_{i-1}(y)$ for all $y\in P$,
				\item if $t_i$ is executed in the forward direction, $\beta\in M_{i-1}(x)$ for some $x\in P$, and $\beta\not\in \effect{t_i}$ then, if $\beta \in \connected(a,M_{i-1}(x))$ and $a\in F(t_i,y)$, then $\beta\in M_{i}(y)$, otherwise $\beta\in M_i(x)$,
				\item if $t_i$ is executed in the reverse direction and $\beta\in \effect{t_i}$ then 
				$\beta\in M_{i-1}(x)$ for some $x\in P$ where $\beta\in F(t_i,x)$, and  $\beta\not\in M_{i}(y)$ for all $y\in P$, and
				\item if $t_i$ is executed in the reverse direction, $\beta\in M_{i-1}(x)$ for some $x\in P$, and $\beta\not\in \effect{t_i}$ then, if $\beta \in \connected(a,M_{i-1}(x))$ and $a\in F(y,t_i)$, then $\beta\in M_{i}(y)$, otherwise $\beta\in M_i(x)$.
			\end{enumerate}
	\end{enumerate}}
\end{proposition}
\paragraph{Proof:}
To begin with, we observe that 
the proofs of clauses (1) and (2)(a) follow directly from clauses (1) and (2) of Propositions~\ref{prop1} 
and~\ref{prop2}. Clause (2)(b) follows from Definition~\ref{forward}(4) and Definition~\ref{forw}. Clause (2)(c) follows from Definition~\ref{forw} and the condition refers to whether
the bond is part of a component manipulated by the forward execution of $t_i$. Similarly,to (2)(a)
clause (2)(d) 
stems from Definition~\ref{br-def}. Finally, Clause (2)(e) follows from Definition~\ref{br-def} and the condition refers to whether
the bond is part of a component manipulated by the reverse execution of $t_i$.
\proofend

In this setting we may establish a loop lemma:
\begin{lemma}[Loop]\label{loopb}{\rm 
		For any forward transition $\state{M}{H}\trans{t}\state{M'}{H'}$ there exists a backward transition
		$\state{M'}{H'} \btrans{t} \state{M}{H}$ and vice versa. 
}\end{lemma}
\paragraph{Proof:}
{Suppose $\state{M}{H}\trans{t}\state{M'}{H'}$. Then $t$ is clearly $bt$-enabled in $H'$. Furthermore,
	$\state{M'}{H'} \btrans{t} \state{M''}{H''}$ where $H'' = H$. In addition, all tokens and bonds
	involved in transition $t$ (except those in $\effect{t}$) will be returned from the outgoing places
	of transition $t$ back to its incoming places. Specifically, for all $a\in A$, it
	is easy to see by the definition of $\btrans{}$  that $a\in M''(x)$ if and only if $a\in M(x)$.
	Similarly,  for all $\beta\in B$,  $\beta\in M''(x)$ if and only if
	$\beta\in M(x)$. The opposite direction can be argued similarly.
}
\proofend

\subsection{Causal-order reversibility}
We now move on to consider causal-order reversibility in RPNs. To define
such as reversible semantics in the presence of cycles, a number of 
issues need to be resolved. To begin with, consider a sequence of transitions
pertaining to the repeated execution of a cycle. Adopting the view that
reversible computation has the ability to rewind \emph{every} executed action of a 
system, we require that each of these transitions is executed in reverse
as many times as it was executed in the forward direction.
Furthermore, the presence of cycles raises questions about the causal relationship 
between transitions of a cycle
as well as of overlapping or even structurally distinct cycles.
In the next subsection we discuss our
adopted notion of transition causality. Subsequently, we develop a theory
for causal-order reversibility in RPNs.

\subsubsection{Causality in cyclic reversing Petri nets}
A cycle in a reversing Petri net is associated with a cyclic path in the net's graph structure. It contains a sequence of transitions
where an outgoing place of the last transition coincides with an incoming place of the first transition.
Note that a cycle in the graph of a reversing Petri net
does not necessarily imply the repeated execution of its transitions since, for instance, entrance
to the cycle may require a token or a bond that has been directed into a different part of the net during execution
of the cycle.

In the standard approach to causality in classical Petri nets~\cite{PNs}, a causal
link is considered to exist between two transitions if one produces
tokens that are used to fire the other.  This relation 
is used to define a ``causal order", $\prec$, which is transitive so that if a transition $t_1$ causally precedes $t_2$ and $t_2$ causally 
precedes $t_3$, then $t_1$ also causally precedes $t_3$.

Adapting this notion in the context of cycle execution, consider  a cycle with
transitions $t_1$ and $t_2$, executed twice yielding the transition instances $t_1^1$, $t_2^1$, $t_1^2$, $t_2^2$, where $t_i^j$
denotes the $j$-th execution of transition $t_i$. Furthermore, suppose that $t_1$ produces tokens that are consumed by $t_2$ and vice versa. This implies the causal order relation
$\prec$, such that $t_1^1\prec t_2^1\prec t_1^2\prec t_2^2$, allowing us to conclude that each execution of the cycle causally
precedes any subsequent executions. This is a natural conclusion in the case of the consecutive execution of cycles,
since a second execution of a cycle cannot be initiated before the first one is completed. This is because the tokens manipulated by the first transition of the cycle need to return to its input places before the transition can be repeated.

\begin{figure}[t]
	\centering
	\subfigure{\includegraphics[width=5.5cm]{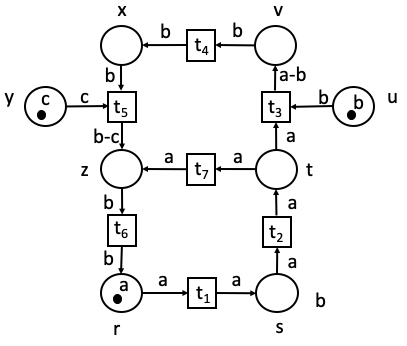}}
	\subfigure{	\includegraphics[width=.55cm]{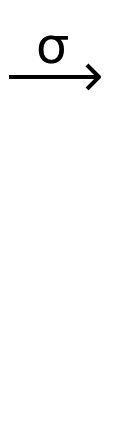}}
	\subfigure{\includegraphics[width=5.5cm]{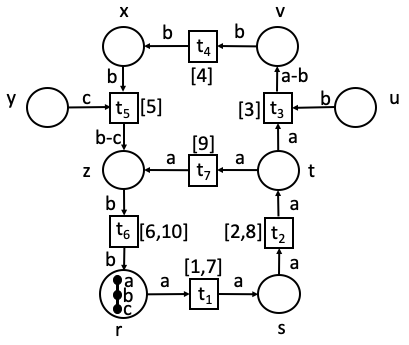}}
	\caption{
		RPN with overlapping cycles $\sigma_1=\langle t_1, t_2, t_3, t_4, t_5, t_6\rangle$ and 
		$\sigma_2 = \langle t_1, t_2, t_7, t_6\rangle$, and the state arising after the forward execution of $\sigma=\sigma_1\sigma_2$}
	\label{cycles1}
\end{figure}

Let us now move on to determining when a token produced by a transition is consumed by
another. In RPNs this concept acquires an additional complexity due to the fact 
that tokens are distinguished by names and the fact that the creation of bonds
between tokens may disguise the causal relation between transitions.
For instance, consider the example of 
Figure~\ref{cycles1}. This RPN features two overlapping cycles, which can be executed sequentially.
Suppose we execute the outer cycle (transition sequence $\langle t_1, t_2, t_3, t_4, t_5, t_6\rangle$) followed
by the inner cycle (transition sequence $\langle t_1, t_2, t_7,t_6\rangle$). 

Observing the token manipulation of the transition
instances as captured by the arcs of the transition, we obtain the order $t_1^1\prec t_2^1\prec t_3^1\prec t_4^1\prec t_5^1\prec t_6^1$ and
$t_1^2\prec t_2^2\prec t_7^1\prec t_6^2$. However by
simply observing the structure of the RPN there is no evidence that $t_1$ consumes tokens
produced by $t_6$. Nonetheless, in this scenario transition instance $t_3^1$ has bonded tokens
$a$ and $b$ and, thus, transition instance $t_1^2$ requires bond $a-b$ to be produced and placed at $r$ by $t_6$ before transition $t_1$ can be executed for the second time. 
Thus, $t_6^1 \prec t_1^2$ also holds.

Note that, if the two cycles were not considered to be causally dependent and were allowed to reverse in any order, then, 
reversal of the first  before the second one would disable the reversal of the second cycle. This is because
reversing transition $t_3$ would return token $b$ to place $u$, thus disabling a second reversal of
transition $t_6$ (and consequently the reversal of the inner cycle).

\begin{figure}[t]
	\centering
	\subfigure{\includegraphics[width=5cm]{cycle20.png}}
	\subfigure{\includegraphics[width=.6cm]{sigma.png}}
	\subfigure{\includegraphics[width=5cm]{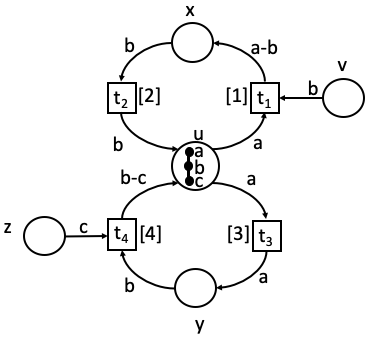}}
	\caption{Causally dependent cycles, where $\sigma=\langle t_1, t_2, t_3,t_4\rangle$ }
	\label{cycles2}
\end{figure}

Similarly, in the example of Figure~\ref{cycles2} we observe two cycles that are structurally independent but
where the presence of common tokens between the two cycles creates a dependence between their executions.
For instance, suppose that the upper cycle is initially selected via execution of transition $t_1$.
This choice disables the lower cycle, which is only re-enabled once the upper cycle is completed and token $a$
is returned to place $u$. As a result, the execution of $t_3$, and thus the lower cycle, following an execution of the upper cycle, is
considered to be causally dependent on the execution of $t_2$. 

The above examples highlight that syntactic token independence between two transitions
or cycles does not preclude their causal dependence. Instead, causal dependence is 
determined by the path that tokens follow: two transition occurrences are causally 
dependent, if a token produced by the one occurrence was subsequently used to
fire the other occurrence.
To capture this type of dependencies, we adopt the following 
definitions.
\begin{definition}\label{transOccurr}{\rm
		Consider a state $\state{M}{H}$ and a transition $t$. We refer to $(t,k)$  as a \emph{transition occurrence} in $\state{M}{H}$ if $k \in H(t)$.
}\end{definition}

\begin{definition}\label{co-dep}{\rm
		Consider  
		a state  $\state{M}{H}$ and suppose $\state{M}{H}\trans{t} \state{M'}{H'}$
		with $(t,k)$, $(t',k')$ transition occurrences in $\state{M'}{H'}$, $k=\max(H(t))$.
		We say that $(t,k)$ \emph{causally depends} on $(t',k')$ denoted by $(t',k')\prec 
		(t,k)$, if $k'<k$ and  there exists $a\in F(x,t)$
		where $\connected(a,M(x))\cap \effects{t'}\neq \emptyset$. 
}\end{definition}

Thus, a transition occurrence $(t,k)$ causally depends on a preceding 
transition occurrence $(t',k')$ if one or more tokens used during the 
firing of $(t,k)$ was produced by $(t',k')$. Note that the tokens employed 
during a transition in a specific marking are determined by the connected 
components of $F(x,t)$ in the marking. 
For example,
in Figure~\ref{cycles1} we have $(t_5,5)\prec (t_7,9)$  and in Figure~\ref{cycles2}  
$(t_1,1)\prec(t_4,4)$, where in each case token $a$ has been transferred from its initial place through $(t_5,5)$ to $(t_7,9)$ and through $(t_1,1)$ to $(t_4,4)$, respectively.

\subsubsection{Causal reversing}
Following this approach to causality, we now move on to define causal-order
reversibility in reversing Petri nets. As expected,
we consider a transition $t$ to be enabled for causal-order reversal only if all
transitions that are causally dependent on it have either been reversed or not 
executed. To this respect, relation $\prec$ becomes an important piece of machinery
and we extend the notion of a \emph{state} for the purposes of causal
dependence to a triple $\cstate{M}{H}{\prec}$ where $\prec$ captures
the causal dependencies that have formed up to the creation of the state.
We assume that in the initial state $\prec = \emptyset$ and
we extend the definition of forward execution as follows:

\begin{definition}{\rm \label{cforw}
		Given a reversing Petri net $(A,P,B,T,F)$, a state $\langle M,H,\prec\rangle$, 
		and a transition $t$ forward-enabled in 
		$\state{M}{H}$, we write $\cstate{M}{H}{\prec}
		\trans{t} \cstate{M'}{H'}{\prec'}$
		where $M'$ and $H'$ are defined as in Definition~\ref{forw}, and
		\[
		\prec' \;= \;\prec\cup \{((t',k'),\!(t,k))\mid 
		k \!=\!\max(H'(t)), (t,k) \mbox{ causally depends on } (t',k') \}
		\]
}\end{definition} 
We may now define that a transition is enabled for causal-order reversal as follows:
\begin{definition}\label{co-enabled}{\rm
		Consider a state $\cstate{M}{H}{\prec}$ 
		and a transition $t\in T$. Then $t$, $H(t)\neq \emptyset$, is $co$-enabled (causal-order reversal
		enabled) in  $\langle M, H,\prec\rangle$ if
		\begin{enumerate}
			\item for all  $x\in t\circ$, if $a\in F(t,x)$ then $a\in M(x)$ and if $\beta\in F(t,x)$ then $\beta\in M(x)$,
			and
			\item there is no transition occurrence $(t',k') \in \langle M, H,\prec\rangle$
			with $(t,k)\prec (t',k')$, for $k=max(H(t))$. 
		\end{enumerate}
}\end{definition}

According to the definition, an executed transition is 
$co$-enabled if all tokens and bonds required for its reversal
(i.e., in $\effects{t}$) are available in its outgoing places
and there are no transitions which depend on it causally. 
Note that the second condition becomes relevant in the presence of cycles since it is possible that, while
more than one transitions simultaneously have available the tokens required for their
reversal, only one of them is $co$-enabled. Such an example can be seen in the final
state of  Figure~\ref{cycles2} and transitions $t_2$ and $t_4$.

Reversing a transition in a causally-respecting manner is implemented similarly to
backtracking, i.e. the tokens are moved from the outgoing places to the incoming places of the transition and all bonds
created by the transition are broken. In addition, the history function is updated in the same manner as in backtracking, where we remove the key of the reversed transition. 
Finally, the 
causal dependence relation removes all references to the reversed
transition occurrence.

\begin{figure}[t]
	\centering
	\hspace{1.5em}
	\subfigure{\includegraphics[width=5cm]{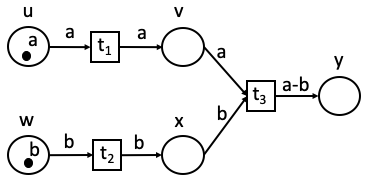}}
	\subfigure{\includegraphics[width=.57cm]{arrow1.png}}
	\subfigure{\includegraphics[width=.55cm]{arrow2.png}}
	\subfigure{\includegraphics[width=.55cm]{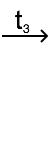}}
	\subfigure{\includegraphics[width=5cm]{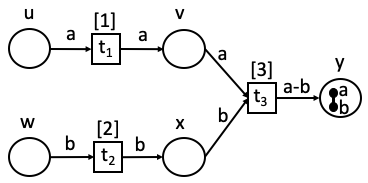}}\\
	\subfigure{\includegraphics[width=.55cm]{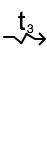}}
	\subfigure{\includegraphics[width=5cm]{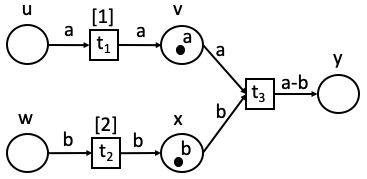}}
	\subfigure{\includegraphics[width=.55cm]{arrow1r.png}}
	\subfigure{\includegraphics[width=5cm]{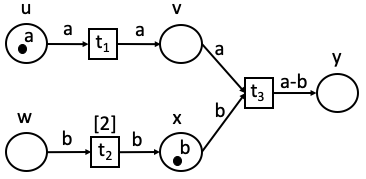}}\\
	\subfigure{\includegraphics[width=.55cm]{arrow2r.png}}
	\subfigure{\includegraphics[width=5cm]{causal6.png}}
	\caption{Causal-order example }\label{co-example}
\end{figure}

\begin{definition}\label{co-def}{\rm
		Given  a state $\langle M, H,\prec\rangle$ and a transition $t$ $co$-enabled in $\langle M, H,\prec\rangle$, we write $\cstate{M}{H}{\prec}
		\ctrans{t} \cstate{M'}{H'}{\prec'}$ for $M'$ and $H'$ as in Definition~\ref{br-def}, and $\prec'$  such that 
		\begin{eqnarray*}
			\prec' &=& \{((t_1,k_1), (t_2,k_2)) \in \prec\; \mid\; k_2\neq k, k=\max(H(t))\}
		\end{eqnarray*}
}\end{definition}

An example of causal-order reversibility can be seen in Figure~\ref{co-example}. Here
we have two independent transitions, $t_1$ and $t_2$ causally preceding transition $t_3$. 
Once the transitions are executed in the order $t_1$, $t_2$, $t_3$, the example 
demonstrates a causally-ordered reversal where $t_3$ is (the only transition that can be) reversed, 
followed by the reversal of its two causes $t_1$ and $t_2$.  In general $t_1$ and $t_2$
can be reversed 
in any order although in the example $t_1$ is reversed before $t_2$. Whenever a transition 
occurrence is reversed its key is eliminated from the history of the transition.

\begin{figure}[t]
	\centering
	\subfigure{\includegraphics[width=5.5cm]{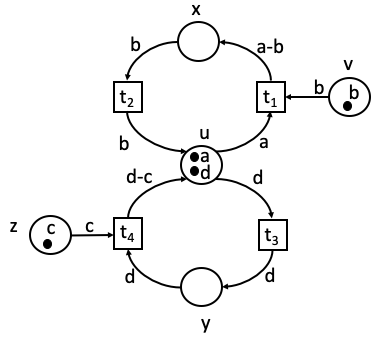}}
	\subfigure{\includegraphics[width=.6cm]{sigmacycle.png}}
	\subfigure{\includegraphics[width=5.5cm]{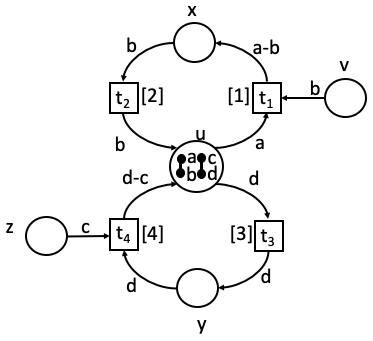}}\\
	
	\subfigure{\includegraphics[width=.6cm]{t3rcycle.png}}
	\subfigure{\includegraphics[width=5.5cm]{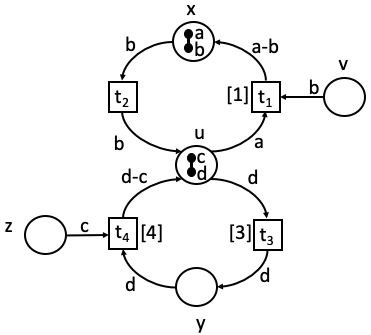}}
	\caption{Causal execution where $\sigma=\langle t_1,t_2,t_3,t_4\rangle$}
	\label{c-cycles}
\end{figure}
As a further example consider the example in Figure~\ref{c-cycles} demonstrating a 
cyclic RPN. Assume that $\sigma=\langle t_1,t_2,t_3,t_4\rangle$, i.e. from the 
initial state of the RPN  the upper cycle is executed followed by the lower 
cycle. The transitions of the two cycles are causally independent since they 
manipulate different sets of tokens and therefore they can be reversed in any 
order. The figure illustrates the reversal of $t_2$ before $t_4$, which returns
the bond between $a\bond b$ to place $x$.

In what follows we write $\fctrans{}$ for $\trans{}\cup\ctrans{}$.  
The following result, similarly to  Proposition~\ref{Prop}, establishes that under the causal-order reversibility semantics, tokens are unique and preserved, bonds are unique, and they can only be created during forward execution and destroyed during reversal.
Note that in what follows we will often omit the causal dependence
relation and simply write $\state{M}{H}$ for states when it is not relevant to
the discussion.
\begin{proposition}\label{Prop4}{\rm Given a reversing Petri net $(A, P,B,T,F)$, an initial state 
		$\langle M_0, H_0\rangle$ and an execution
		$\state{M_0}{H_0} \fctrans{t_1}\state{M_1}{H_1} \fctrans{t_2}\ldots \fctrans{t_n}\state{M_n}{H_n}$, the following hold:
		\begin{enumerate}
			\item For all $a\in A$  and $i$, $0\leq i \leq n$,   $|\{x\in P \mid a\in M_i (x)\}| = 1$.
			\item For all $\beta \in B$ and $i$,  $0\leq i \leq n$, 
			\begin{enumerate}
				\item $0 \leq |\{x \in P \mid \beta\in M_i(x)\}| \leq 1$,
				\item if $t_i$ is executed in the forward direction and $\beta\in \effect{t_i}$, then 
				$\beta\in M_{i}(x)$ for some $x\in P$ where $\beta\in F(t_i,x)$, and  $\beta\not\in M_{i-1}(y)$ for all $y\in P$,
				\item if $t_i$ is executed in the forward direction, $\beta\in M_{i-1}(x)$ for some $x\in P$, and $\beta\not\in \effect{t_i}$, then, if $\beta \in \connected(a,M_{i-1}(x))$ and $a\in F(t_i,y)$ then $\beta\in M_{i}(y)$, otherwise $\beta\in M_i(x)$
				\item if $t_i$ is executed in the reverse direction and $\beta\in \effect{t_i}$, then 
				$\beta\in M_{i-1}(x)$ for some $x\in P$ where $\beta\in F(t_i,x)$, and  $\beta\not\in M_{i}(y)$ for all $y\in P$, and
				\item if $t_i$ is executed in the reverse direction, $\beta\in M_{i-1}(x)$ for some $x\in P$, and $\beta\not\in \effect{t_i}$, then, if $\beta \in \connected(a,M_{i-1}(x))$ and $a\in F(y,t_i)$ then $\beta\in M_{i}(y)$, otherwise $\beta\in M_i(x)$.
			\end{enumerate}
	\end{enumerate}}
\end{proposition}

\paragraph{Proof:} The proof follows along the same lines as that of Proposition~\ref{Prop} with $\btrans{}$ replaced
by $\ctrans{}$.
\proofend

We may now establish the causal consistency of our semantics. 
First we define 
some auxiliary notions. Given a transition $\state{M}{H}\fctrans{t}\state{M'}{H'}$,
we say that the \emph{action} of the transition is
$t$ if $\state{M}{H}\trans{t}\state{M'}{H'}$ and $\underline{t}$ 
if $\state{M}{H}\ctrans{t}\state{M'}{H'}$
and we may write $\state{M}{H}\fctrans{\underline{t}}\state{M'}{H'}$. 
We use $\alpha$ to
range over $\{t,\underline{t} \mid t\in T\}$ and write 
$\underline{\underline{\alpha}} = \alpha$. We extend this
notion to sequences of transitions and, given an execution 
$\state{M_0}{H_0}\fctrans{t_1}\ldots 
\fctrans{t_n}\state{M_n}{H_n}$, we say that the \emph{trace} 
of the execution is
$\sigma=\langle \alpha_1,\alpha_2,\ldots,\alpha_n\rangle$, 
where $\alpha_i$ is the action of transition 
$\state{M_{i-1}}{H_{i-1}}\fctrans{t_i}\state{M_i}{H_i}$,  and write
$\state{M}{H}\fctrans{\sigma}\state{M_n}{H_n}$. Given $\sigma_1 = 
\langle \alpha_1,\ldots,\alpha_k\rangle$, $\sigma_2 = 
\langle \alpha_{k+1},\ldots,\alpha_n\rangle$,
we write $\sigma_1;\sigma_2$ for $\langle \alpha_1,\ldots,\alpha_n\rangle$. 
We may also use the notation $\sigma_1;\sigma_2$ when 
$\sigma_1$ or $\sigma_2$ is a single transition.

An execution of a Petri net can be partitioned
as a set of independent flows of execution
running through the net. We capture these flows by the notion of
causal paths:
\begin{definition}\label{co-path}{\rm
		Given a state $\cstate{M}{H}{\prec}$ and transition occurrences
		$(t_i,k_i)$ in $\cstate{M}{H}{\prec}$, $1\leq i \leq n$, we say that
		$(t_1,k_1),\ldots,(t_n,k_n)$ is a \emph{causal path} in $\cstate{M}{H}{\prec}$,
		if $(t_i,k_i)\prec (t_{i+1}, k_{i+1})$, for all $0\leq i < n$.
}\end{definition}
\begin{figure}[t]
	\centering
	\subfigure{\includegraphics[width=5.5cm]{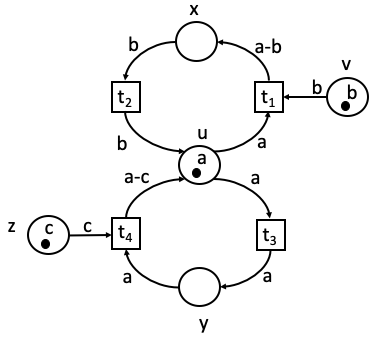}}
	\subfigure{\includegraphics[width=.8cm]{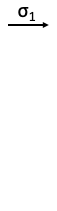}}
	\subfigure{\includegraphics[width=5.5cm]{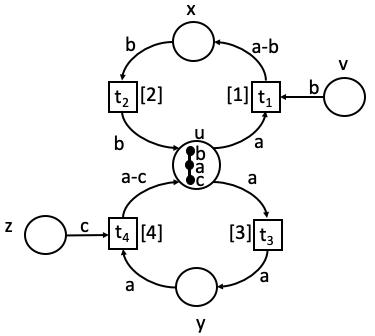}}\\
	\subfigure{\includegraphics[width=5.5cm]{path2.png}}
	\subfigure{\includegraphics[width=.8cm]{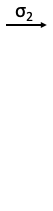}}
	\subfigure{\includegraphics[width=5.5cm]{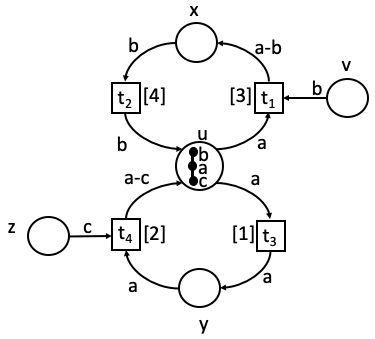}}
	\caption{Causal paths in the context of independent cycles, where $\sigma_1=\langle t_1,t_2,t_3,t_4\rangle$ and $\sigma_2=\langle t_3,t_4,t_1,t_2\rangle$}
	\label{causalPaths1}
\end{figure}

As an example, consider the RPN in Figure~\ref{causalPaths1}
where we denote the first execution by $\cstate{M_0}{H_0}{\emptyset} \trans{\sigma_1}\cstate{M_4}{H_4}{\prec}$ 
for  $\sigma_1=\langle t_1,t_2,t_3,t_4\rangle$, 
and the second execution by $\cstate{M_0}{H_0}{\emptyset}
\trans{\sigma_2}\cstate{M_4'}{H_4'}{\prec'}$
for  $\sigma_2=\langle t_3,t_4,t_1,t_2\rangle$. 
In the case of
$\sigma_1$ we have  $\prec$ to be the transitive
closure of $\{((t_1,1), (t_2,2)),$ $  ((t_2,2),(t_3,3)),((t_3,3),(t_4,4))\}$, 
which results in the causal path  
$(t_1,1),(t_2,2),$ $(t_3,3),(t_4,4)$. In the case of
$\sigma_2$ where the cycles are executed in the opposite order,
$\prec'$ is the transitive
closure of $\{((t_3,1),(t_4,2)),  ((t_4,2),(t_1,3)), ((t_1,3),$ $(t_2,4))\}$, 
and the corresponding causal path is $(t_3,1),(t_4,2),(t_1,3),(t_2,4)$.

\begin{figure}[t]
	\centering
	\subfigure{\includegraphics[width=5.5cm]{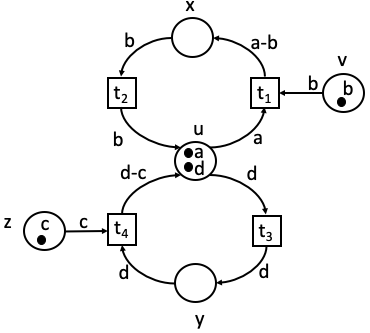}}
	\subfigure{\includegraphics[width=.8cm]{sigma1.png}}
	\subfigure{\includegraphics[width=5.5cm]{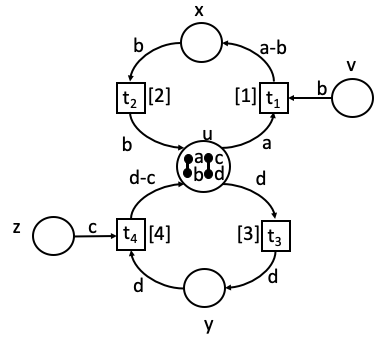}}\\
	\subfigure{\includegraphics[width=5.5cm]{path5.png}}
	\subfigure{\includegraphics[width=.8cm]{sigma2.png}}
	\subfigure{\includegraphics[width=5.5cm]{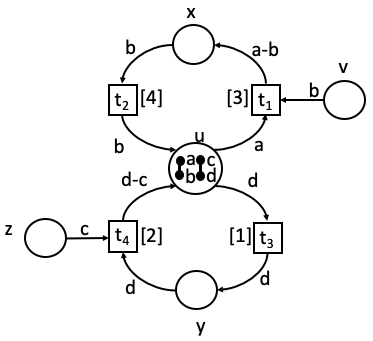}}
	\caption{Causal paths in the context of independent cycles, where $\sigma_1=\langle t_1,t_2,t_3,t_4\rangle$ and $\sigma_2=\langle t_3,t_4,t_1,t_2\rangle$}
	\label{causalPaths2}
\end{figure}

This comes in contrast to the RPN of Figure~\ref{causalPaths2}, which 
contains two independent cycles. 
Here, the  causal dependencies of the first execution (trace $\sigma_1$)
are constructed as 
$(t_1,1)\prec (t_2,2)$ and $(t_3,3) \prec (t_4,4)$, which results in the two 
independent causal paths $\langle (t_1,1),(t_2,2)\rangle$ and 
$\langle (t_3,3),(t_4,4)\rangle$.
Similarly, after execution of $\sigma_2$, 
the causal dependencies are $(t_1,3)\prec (t_2,4)$ and 
$(t_3,1) \prec (t_4,2)$, which results in the causal paths 
$\langle (t_1,3),(t_2,4)\rangle$ and $\langle (t_3,1),(t_4,2)\rangle$. 

As seen from the examples in Figures~\ref{causalPaths1} and~\ref{causalPaths2}, 
the causal paths of an execution capture its causal behavior. 
Based on this concept, we define the notion of causal equivalence for histories by 
requiring that two histories $H$ and $H'$ are causally equivalent if and only
if they contain the same causal paths: 

\begin{definition}\label{eq}{\rm Consider a reversing Petri net $(A,P,B,T,F)$ and two executions
		$\cstate{M}{H}{\prec} \fctrans{\sigma} \cstate{M'}{H'}{\prec'}$ and
		$\cstate{M}{H}{\prec}\fctrans{\sigma'} \cstate{M''}{H''}{\prec''}$. Then the histories
		$H'$ and $H''$ are \emph{causally equivalent}, denoted by $H'\asymp H''$, if
		for each causal path   $(t_1,k_1),\ldots,(t_n,k_n)$ 
		in $\cstate{M'}{H'}{\prec'}$, there is a causal path $(t_1,k_1'),\ldots,
		(t_n,k_n')$ in $\cstate{M''}{H''}{\prec''}$, and vice versa.

		We extend this notion and write $\cstate{M}{H}{\prec}\asymp\cstate{M'}{H'}{\prec'}$ 
		if and only if $M=M'$ and $H\asymp H'$.
}\end{definition}
Returning to the example in Figure~\ref{causalPaths1} we observe that while the two executions result in the same marking,
the resulting states do not have the same causal paths and, as such,
they are not considered as causally equivalent.

We may now establish the Loop lemma. 
\begin{lemma}[Loop]\label{loopc}{\rm 
		For any forward transition $\state{M}{H}\trans{t}\state{M'}{H'}$ there exists a backward transition
		$\state{M'}{H'} \ctrans{t} \state{M}{H}$ and for any backward transition $\state{M}{H} \ctrans{t} \state{M'}{H'}$ there exists a forward transition $\state{M'}{H'}\trans{t}\state{M}{H''}$ where $H\asymp H''$.
}\end{lemma}

\paragraph{Proof:} The proof of the first direction follows 
along the same lines as that of Lemma~\ref{loopb} with $\btrans{}$
replaced by $\ctrans{}$. For the other direction suppose $\state{M}{H}\ctrans{t}\state{M'}{H'}\trans{t}\state{M}{H''}$. 
To begin with, we may observe that, as with Lemma~\ref{loopb}, $M=M''$.
To show that $H\asymp H''$, we observe that $H=H''$ with the exception of
$t$, where, if $k=\max(H(t))$,
and $k'=\max(\{0\}\cup\{k''|(t',k'')\in H'(t'), t'\in T\})+1$, then
$H''(t) = (H(t)-\{k\})\cup\{k'\}) $.
Furthermore, since $t$ is $co$-enabled in $\state{M}{H}$, $(t,k)$ must
be the last transition occurrence in all the causal paths it occurs in,
and we may observe that $H''$ contains the same causal paths with 
$(t,k)$ replaced by $(t,k')$.
As a result it must be that $H\asymp H''$ and the result follows.
\proofend

We now proceed to define causal equivalence on traces, a notion that employs the concept of concurrent transitions:
\begin{definition}{\rm
		Actions  $\alpha_1$ and $\alpha_2$ are \emph{concurrent} in state $\cstate{M}{H}{\prec}$, if whenever
		$\cstate{M}{H}{\prec}\fctrans{\alpha_1}\cstate{M_1}{H_1}{\prec_1}$ and $\cstate{M}{H}{\prec}\fctrans{\alpha_2}\cstate{M_2}{H_2}{\prec_2}$
		then $\cstate{M_1}{H_1}{\prec_1}\fctrans{\alpha_2}\cstate{M'}{H'}{\prec'}$ and $\cstate{M_2}{H_2}{\prec_2}\fctrans{\alpha_1}$ $\cstate{M''}{H''}{\prec''}$,
		where $\cstate{M'}{H'}{\prec'} \asymp \cstate{M''}{H''}{\prec''}$.
}\end{definition}

Thus, two actions are concurrent if execution of the one does not preclude the other and the two execution 
orderings lead to 
causally equivalent states. The condition on final states being
equivalent is required to rule out transitions constituting self-loops
to/from the same place that are causally dependent on each other.
\begin{definition}\label{co-executions}{\rm  \emph{Causal equivalence on traces}, denoted by $\asymp$, is the least
		equivalence relation closed under composition of traces such that 
		(i) if $\alpha_1$ and $\alpha_2$ are concurrent actions then $\alpha_1;\alpha_2\asymp \alpha_2;\alpha_1$ and
		(ii) $\alpha ; \underline{\alpha} \asymp \epsilon$.
}\end{definition}

The first clause states that in two causally-equivalent traces concurrent actions
may occur in any order and the second clause states that it is possible to ignore transitions that have occurred in both
the forward and the reverse direction.

The following proposition establishes that two transition instances belonging
to distinct causal paths are in fact concurrent transitions and thus can be executed in any order. 

\begin{proposition}\label{conc-transitions}{\rm
		Consider a reversing Petri net $(A,P,B,T,F)$ and suppose $\langle M, H,{\prec}\rangle
		\trans{t_1} \cstate{M_1}{H_1}{\prec_1}\trans{t_2} \cstate{M_2}{H_2}{\prec_2}$,
		where the executions of $t_1$ and $t_2$ correspond to transition instances $(t_1,k_1)$ and  $(t_2,k_2)$ in $\cstate{M_2}{H_2}{\prec_2}$. If
		there is no causal path $\pi$ in $\cstate{M_2}{H_2}{\prec_2}$ with $(t_1,k_1)\in\pi$ and $(t_2,k_2)\in\pi$,
		then
		$(t_1,k_1)$ and $(t_2,k_2)$ are concurrent transition occurrences in $\cstate{M}{H}{\prec}$. 
}\end{proposition}

\paragraph{Proof:}
Since there is no causal path containing both $(t_1,k_1)$ and $(t_2,k_2)$ in $\cstate{M_2}{H_2}{\prec_2}$,
we conclude that $(t_1,k_1)\not\prec_2 (t_2,k_2)$.
This implies that the two transition occurrences do not handle any common tokens and they can
be executed in any order leading to the same marking. Thus, they are 
concurrent in $\cstate{M}{H}{\prec}$.
\proofend

We note that causally-equivalent states can execute the same transitions.
\begin{proposition}{\rm
		Consider a reversing Petri net $(A,P,B,T,F)$ and states $\langle M, H_1,{\prec_1}\rangle\asymp\langle M, H_2,{\prec_2}\rangle$. 
		Then $\cstate{M}{H_1}{\prec_1}\fctrans{\alpha}\cstate{M_1}{H_1'}{\prec_1'}$ if and only if  
		$\cstate{M}{H_2}{\prec_2}\fctrans{\alpha}\cstate{M_2}{H_2'}{\prec_2'}$, where $\langle M_1, H_1',{\prec_1'}\rangle\asymp\langle M_2, H_2',{\prec_2'}\rangle$. 
		\label{extendStates}
}\end{proposition}
\paragraph{Proof:}
It is easy to see that if a transition $\alpha$ is enabled in $\cstate{M}{H_1}{\prec_1}$
it is also enabled in $\cstate{M}{H_2}{\prec_2}$. Therefore, if  $\cstate{M}{H_1}{\prec_1}\fctrans{\alpha}\cstate{M_1}{H_1'}{\prec_1'}$
then $\cstate{M}{H_2}{\prec_2}\fctrans{\alpha}\cstate{M_2}{H_2'}{\prec_2'}$ where
$M_1=M_2$, and vice versa. In order
to show that $H_1'\asymp H_2'$ two cases exist:
\begin{itemize}
	\item Suppose $\alpha$ is a forward transition corresponding to transition
	occurrence $(t,k_1)$ in $\cstate{M_1}{H_1'}{\prec_1'}$ and transition occurrence $(t,k_2)$ in $\langle; _2, H_2',$ $\prec_2'\rangle$.
	Suppose that $(t',k_1')\prec_1' (t,k_1)$.
	Then, $\effects{t'}\cap \connected(a,M(x))\neq \emptyset$
	for some $a\in F(x,t)$. Since $H_1\asymp H_2$, this implies that $(t',k_2')\prec_2' (t,k_2)$
	where $k_2' = \max(H_2(t'))$. 
	since $H_1\asymp H_2$.)
	Therefore, for all causal paths $\pi$ in $\cstate{M}{H_1}{\prec_1}$,
	if the last transition occurrence of $\pi$ causes $(t,k_1)$ then
	$\pi;(t,k_1)$ is a causal path of $\cstate{M_1}{H_1'}{\prec_1'}$ and, if not, then $\pi$ is
	a causal path in $\cstate{M_1}{H_1'}{\prec_1'}$. The same holds for causal paths
	in $\cstate{M_2}{H_2'}{\prec_2'}$ and $(t,k_2)$. Consequently, we deduce that $H_1'\asymp H_2'$, as required.
	\item Suppose that $\alpha$ is a reverse transition, i.e. $\alpha = \underline{t}$ 
	for some $t$, and consider the causal paths of
	$H_1'$ and $H_2'$. Since $\alpha$ is a reverse transition, there exists no transition occurrence
	caused by $(t,\max(H_1(t)))$ in $\cstate{M}{H_1}{\prec_1}$ 
	and no transition occurrence caused by $(t,\max(H_2(t)))$ in $\cstate{M}{H_2}{\prec_2}$. 
	As such, $(t,\max(H_1(t)))$ and  $(t,\max(H_2(t)))$ are the last
	transition occurrences in all paths in $\cstate{M}{H_1}{\prec_1}$ and $\cstate{M}{H_2}{\prec_2}$,
	respectively, in which they belong.
	Reversing the transition occurrences results in their elimination from these causal paths. 
	Therefore, we observe that for each causal path in $\cstate{M_1}{H_1'}{\prec_1'}$ there is
	an equivalent causal path in $\cstate{M_2}{H_2'}{\prec_2'}$, and vice versa. 
	Thus $H_1'\asymp H_2'$ as required.
	\proofend
\end{itemize}

Note that the above result can be extended to sequences of transitions:
\begin{corollary}\label{conc-paths}{\rm
		Consider a reversing Petri net $(A,P,B,T,F)$ and states $\langle M, $ $H_1,{\prec_1}\rangle\asymp\langle M, H_2,{\prec_2}\rangle$. 
		Then $\cstate{M}{H_1}{\prec_1}\fctrans{\sigma}\cstate{M_1}{H_1'}{\prec_1'}$ if and only if  
		$\cstate{M}{H_2}{\prec_2}\fctrans{\sigma}\cstate{M_2}{H_2'}{\prec_2'}$, where $\langle M_1, H_1',{\prec_1'}\rangle\asymp\langle M_2, H_2',{\prec_2'}\rangle$.  
}\end{corollary}

\remove{We may also prove that any two simultaneously enabled reverse transitions are in
	fact concurrent.
	\begin{proposition}\label{rev-conc-transitions}{\rm
			Suppose $\langle M, H,{\prec}\rangle \ctrans{t_1} \cstate{M_1}{H_1}{\prec_1}$ 
			and $\langle M, H,{\prec}\rangle \ctrans{t_2} \cstate{M_2}{H_2}{\prec_2}$.
			Then $\underline{t_1}$ and $\underline{t_2}$ are concurrent transitions in 
			$\cstate{M}{H}{\prec}$. 
	}\end{proposition}
	
	\paragraph{Proof:}
	Suppose that $\langle M, H,{\prec}\rangle \ctrans{t_1} \cstate{M_1}{H_1}{\prec_1}$ 
	and $\langle M, H,{\prec}\rangle \ctrans{t_2} \cstate{M_2}{H_2}{\prec_2}$.
	...
	\proofend
}

The main result, Theorem~\ref{main} below, states  that  two computations beginning in the same initial state lead to equivalent
states if and only if the sequences of executed transitions of the two computations are causally equivalent. 
This guarantees the consistency of the approach since reversing transitions in causal order is in a sense equivalent
to not executing the transitions in the first place. Reversal does not give rise
to previously unreachable states, on the contrary, it gives rise to exactly the same markings and causally-equivalent
histories due to the different keys being possibly assigned because of the different ordering of transitions.

\begin{theorem}\label{main}{\rm Consider executions $\state{M}{H} \fctrans{\sigma_1} \state{M_1}{H_1}$ and $\state{M}{H}\fctrans{\sigma_2} \state{M_2}{H_2}$. Then, $\sigma_1\asymp\sigma_2$ if and only if   $\state{M_1}{H_1}\asymp\state{M_2}{H_2}$.
	}
\end{theorem}

For the proof of Theorem~\ref{main} we employ some intermediate results. To begin with, the lemma below
states that causal equivalence allows the permutation of reverse and forward transitions that have no causal relations between them. 
Therefore, computations are allowed to reach for the maximum freedom of choice going backward and then continue forward.

\begin{lemma}\label{perm}{\rm \
		Let $\sigma$ be a trace. Then there exist traces $r,r'$ both forward such that  
		$\sigma\asymp\underline{r};r'$ and if $\state{M}{H} \frtrans{\sigma}\state{M'}{H'}$ then $\state{M}{H} \frtrans{\underline{r};r'}\state{M'}{H''}$, where $H'\asymp H''$.
}\end{lemma}

\paragraph{Proof:}
We prove this by induction on the length of  $\sigma$ and the
distance from the beginning of $\sigma$ to the earliest pair of transitions
that contradicts the property $\underline{r};r'$. If there is no such
contradicting pair then the property is trivially satisfied.  If not, we distinguish the following cases:
\begin{enumerate}
	\item If the first contradicting pair is of the form $t;\underline{t}$
	then we have $\state{M}{H}\fctrans{\sigma_1}\state{M_1}{H_1}\fctrans{t}\state{M_2}{H_2}\fctrans{\underline{t}}\state{M_3}{H_3}\fctrans{\sigma_2}\state{M'}{H'}$ where $\sigma=\sigma_1;t;\underline{t};\sigma_2$. By the Loop Lemma $\state{M_1}{H_1}=\state{M_3}{H_3}$, which yields $\state{M}{H} \fctrans{\sigma_1} \state{M_1}{H_1} \fctrans{\sigma_2} \state{M'}{H'}$. Thus we may remove the two transitions from
	the sequence, the length of $\sigma$ decreases, and the proof follows
	by induction.
	\item If the first contradicting pair is of the form $t;\underline{t}'$ then
	we observe that the specific occurrences of $t$ and $\underline{t}'$ must be
	concurrent. Specifically we have $ \state{M}{H} 
	\fctrans{\sigma_1}\state{M_1}{H_1}\fctrans{t}\state{M_2}{H_2}\fctrans{\underline{t'}}
	\state{M_3}{H_3}\fctrans{\sigma_2}\state{M'}{H'}$ where $\sigma=\sigma_1 ; t
	;\underline{t'}; \sigma_2$. Since
	action $t'$ is being reversed, all 
	transition occurrences that are causally dependent on it have either
	not been executed up to this point or they have already been reversed. This implies
	that in $\state{M_2}{H_2}$ it was not the case that $(t',max(H_2(t'))$ was causally dependent on $(t,max(H_2(t))$.
	As such, by Proposition~\ref{conc-transitions}, $\underline{t'}$ and $t$ are concurrent transitions
	and $t'$ can be reversed before
	the execution of $t$
	to yield $ \state{M}{H} \fctrans{\sigma_1}\state{M_1}{H_1}\fctrans{\underline{t'}}\state{M_2'}{H_2'}
	\fctrans{t}\state{M_3}{H_3'}\fctrans{\sigma_2}\state{M'}{H''}$, where $H_3' \asymp H_3$ and $H'\asymp H''$. This results in a
	later earliest contradicting pair and by induction the result follows.
	\proofend
\end{enumerate}

From the above lemma we conclude the following corollary establishing that causal-order
reversibility is consistent with standard forward execution in the sense that causal execution will not generate
states that are unreachable in forward execution:

\begin{corollary}\label{equivalent-executions}{\rm\ \
		Suppose that  $H_0$ is the initial history. If $\state{M_0}{H_0} \fctrans{\sigma} \state{M}{H}$, and $\sigma$ is a
		trace with both forward and backward transitions then
		there exists a transition $\state{M_0}{H_0}\fctrans{\sigma'}\state{M}{H'}$, where $H\asymp H'$ 
		and $\sigma'$ a trace of forward transitions.
}\end{corollary}
%
\paragraph{Proof:} According to
Lemma~\ref{perm}, $\sigma\asymp \underline{r};r'$ where both $r$ and $r'$ are forward
traces. Since, however, $H_0$ is the initial history it must be that $r$ is empty. This
implies that $\state{M_0}{H_0}\fctrans{r'}\state{M}{H'}$, $H\asymp H'$
and $r'$ is a 
forward trace. Consequently, writing $\sigma'$ for $r'$, the result follows.
\proofend

\begin{lemma}\label{short}{\rm\ \
		Suppose $\state{M}{H}\fctrans{\sigma_1}\state{M'}{H_1}$ and
		$\state{M}{H}\fctrans{\sigma_2}\state{M'}{H_2}$, where $H_1\asymp H_2$ and
		$\sigma_2$ is a forward trace. Then, there exists a forward trace
		$\sigma_1'$ such that $\sigma_1 \asymp \sigma_1'$.
}\end{lemma}

\paragraph{Proof:}
If  $\sigma_1$ is forward then $\sigma_1 = \sigma_1'$ and the result follows
trivially. Otherwise, we may prove the lemma by induction on the length of
$\sigma_1$.
We begin by noting that, by Lemma~\ref{perm},
$\sigma_1\asymp\underline{r};r'$ and $\state{M}{H}\fctrans{\underline{r};r'}
\state{M'}{H_1}$. 
Let  $\underline{t}$ be
the last action in $\underline{r}$. 
Given that $\sigma_2$ is a forward execution that 
simulates $\sigma_1$, it must be that $r'$ contains a forward execution 
of transition $t$ so that $\state{M'}{H_1}$ and $\state{M'}{H_2}$ contain the same causal 
paths involving transition $t$ (if not we would have 
$|H_1(t)|<|H_2(t)|$ leading to a contradiction). 
Consider the earliest  occurrence of $t$ in $r'$. If $t$ is the first
transition in $r'$, by the Loop Lemma we may remove the pair of opposite transitions
and the result follows by induction. Otherwise, suppose 
$\state{M}{H}\fctrans{\underline{r_1}}\fctrans{\underline{t}}
\fctrans{{r_1'}} \state{M_1}{H_3} \fctrans{t*}\fctrans{t}\state{M_1'}{H_4}
\fctrans{r_2'}\state{M'}{H_1}$, where $r = r_1;t$ and $r'=r_1';t*;t;r_2$.
Two cases exist:
\begin{enumerate}
	\item Suppose $t*\in\sigma_2$. Let us denote by $num(t,\sigma)$, 
	the number of executions of transition $t$ in a sequence of
	transitions $\sigma$. We observe that since $\sigma_2$ contains no
	reverse executions of $t$, it must be that $num(t,r') = num(t,\sigma_2) + num(t,r)$.
	Suppose that the transition occurrences of $t*$ and $t$ as shown in the execution
	belong to a common causal path. We may extend this path with the succeeding 
	occurrences of $t$ and obtain a causal path such that $t*$ is succeeded by
	$num(t,\sigma_2) + num(t,r)$ occurrences of $t$. We observe that it is impossible to
	obtain such a causal path in $\state{M'}{H_2}$, since $t*$ is followed by fewer occurrences of $t$ in $\sigma_2$.
	This contradicts the assumption that $H_1\asymp H_2$. We conclude that the transition
	occurrences of $t$ and $t*$ above do not belong to any common causal path and
	therefore, by Proposition~\ref{conc-transitions}, the two transition occurrences are
	concurrent in $\state{M_1}{H_3}$.
	\item
	Now suppose that $t*\not\in \sigma_2$. Since $H_1(t*)\neq \emptyset$ it must
	be that $H_2(t*)\neq \emptyset$ and $|H(t*)| = |H_1(t*)|= |H_2(t*)|$. As such, it
	must be that $t*\in r$ and that its reversal has preceded the reversal of $t$.
	Let us suppose that the transition occurrences of $t*$ and $t$ as shown in the execution
	belong to a common causal path. This implies that a causal path
	with $t*$ preceding $t$ also occurs in $H_2$ as well as in $H$. If we
	observe that $t*$ has reversed before $t$ we conclude that  $t*$ does not cause the preceding occurrence of $t$. As such
	there is no causal path within $\state{M}{H}$ or $\state{M'}{H_2}$ containing both $t$ and $t*$, which results in a contradiction. We
	conclude that the forward occurrences of $t$ and $t*$ are,  by Proposition~\ref{conc-transitions}, concurrent in $\state{M_1}{H_3}$.
\end{enumerate}
Given the above, since the occurrences of $t$ and $t*$ are concurrent the two occurrences may be swapped to yield
$\state{M}{H}\fctrans{\underline{r_1}}\fctrans{\underline{t}}
\fctrans{{r_1'}}  
\state{M_1}{H_3} \fctrans{t}\fctrans{t*}\state{M_1'}{H_4'}\fctrans{r_2'}\state{M'}{H_1'}$ where $H_4\asymp H_4'$ and, by Corollary~\ref{conc-paths},  $H_1\asymp H_1'$.
By repeating the process for the remaining transition occurrences in $r_1'$, this implies
that we may permute $t$ with transitions in $r_1'$ to yield the sequence $\underline{t};t$. By the Loop Lemma we may remove the pair of opposite
transitions and obtain a shorter equivalent trace, also
equivalent to $\sigma_2$ and conclude by induction.
\proofend

We now proceed with the proof of Theorem~\ref{main}:

\paragraph{Proof of Theorem~\ref{main}:}
Suppose $\state{M}{H} \fctrans{\sigma_1} \state{M_1}{H_1}$, $\state{M}{H} \fctrans{\sigma_2} \state{M_2}{H_2}$
with  $\state{M_1}{H_1}\asymp\state{M_2}{H_2}$. 
We prove that $\sigma_1\asymp \sigma_2$ by using a lexicographic induction on the pair consisting 
of the sum of the lengths of $\sigma_1$ and $\sigma_2$ and the depth of the earliest disagreement
between them. By Lemma~\ref{perm} we may suppose that $\sigma_1$ and $\sigma_2$
satisfy the property
$\underline{r};r'$. Call $t_1$ and $t_2$ the earliest actions where they disagree. There are three
cases in the argument depending on whether these are forward or backward.
\begin{enumerate}
	
	\item If $t_1$ is backward and $t_2$ is  forward, we have $\sigma_1=\underline{r};\underline{t_1};u$ 
	and $\sigma_2=\underline{r};t_2;v$ for some $r,u,v$. Lemma~\ref{short} applies to $t_2;v$, 
	which is forward, and $\underline{t_1};u$, which contains both forward and backward actions and thus,
	by the lemma, it  has a shorter forward equivalent. Thus, $\sigma_1$ has a shorter forward 
	equivalent and the result follows by induction.
	
	\item If $t_1$ and $t_2$ are both forward  then it must be the
	case that $\sigma_1 = {\underline{r};r'};t_1; u$
	and $\sigma_2 =  {\underline{r};r'}; t_2; v$, for some $r$, $u$, $v$. Note that
	it must be that $t_1\in v$ and $t_2\in u$. 
	If not, we would have $|H_1(t_1)|\neq |H_2(t_1)|$, and similarly for
	$t_2$, which contradicts the assumption that ${H_1}\asymp {H_2}$.
	As such, we may write $\sigma_1 =  {\underline{r};r'};t_1;u_1;t_2;u_2$, 
	where $u=u_1;t_2;u_2$
	and $t_2$ is the first occurrence of $t_2$ in $u$. Consider $t*$ the 
	action immediately preceding $t_2$. We observe that $t*$ and $t_2$ 
	cannot belong to
	a common causal path in $\state{M_1}{H_1}$, since an equivalent causal path is 
	impossible to
	exist in $\state{M_2}{H_2}$. This is due to the assumption that $\sigma_1$ 
	and $\sigma_2$ coincide up to transition sequence  {$\underline{r};r'$}. 
	Thus, we conclude by  Proposition~\ref{conc-transitions} that $t*$ and 
	$t_2$ are in fact concurrent and can be swapped.
	The same reasoning may be used
	for all transitions preceding $t_2$ up to and including
	$t_1$, which leads to the conclusion that
	$\sigma_1\asymp  {\underline{r};r'};t_2;t_1; u_1;u_2$. This results in an 
	equivalent execution of the same length with a later earliest divergence with 
	$\sigma_2$ 
	and the result follows by the induction hypothesis.
	
	\item If $t_1$ and $t_2$ are both backward, we have $\sigma_1=\underline{r};\underline{t_1};u$ 
	and $\sigma_2=\underline{r};\underline{t_2};v$ for some $r,u,v$. Two cases exist:
	\begin{enumerate}
		\item If $\underline{t_1}$ occurs in $v$, then we have that $\sigma_2=\underline{r};\underline{t_2};\underline{v_1};\underline{t_1};v_2$.
		Given that $t_1$ reverses right after $\underline{r}$ in
		$\sigma_1$, we may conclude that there is no transition occurrence
		at this point that causally depends on $t_1$. As such it
		cannot have caused the transition occurrences of $t_2$ and
		${v_1}$ whose reversal precedes it in $\sigma_2$. 
		This implies that the reversal of $t_1$
		may be swapped in $\sigma_2$ with each of the preceding
		transitions, to give
		$\sigma_2\asymp\underline{r};\underline{t_1};\underline{t_2};\underline{v_1};v_2$.
		This results in an equivalent execution of the same length with a later earliest divergence 
		with $\sigma_1$ and the result follows by the induction hypothesis.
		\item If $\underline{t_1}$ does not occur in $v$, this implies that $t_1$ occurs
		in the forward direction in $u$, i.e. $\sigma_1=\underline{r};\underline{t_1};u_1;t_1;u_2$, where $u = u_1;t_1;u_2$ with
		the specific occurrence of $t_1$ being the first such occurrence in $u$. 
		Using similar arguments as those in Lemma~\ref{short},
		we conclude that $\sigma_1\asymp\underline{r};\underline{t_1};t_1;u_1;u_2
		\asymp \underline{r};u_1;u_2$, an equivalent execution of shorter length for $\sigma_1$ and the result follows by the induction hypothesis.
	\end{enumerate}

	We may now prove the opposite direction. Suppose that $\sigma_1 \asymp \sigma_2$
	and $\state{M}{H}\fctrans{\sigma_1}\state{M_1}{H_1}$ and $\state{M}{H}\fctrans{\sigma_2}\state{M_2}{H_2}$. 
	We will show that $\state{M_1}{H_1}\asymp \state{M_2}{H_2}$.
	The proof is by induction on the number of rules, $k$, applied to establish 
	the equivalence $\sigma_1 \asymp \sigma_2$.
	For the base case we have $k=0$, which implies that $\sigma_1=\sigma_2$
	and the result trivially follows. For the inductive step, let us assume that
	$\sigma_1\asymp \sigma_1'\asymp \sigma_2$, where $\sigma_1$ can be transformed
	to $\sigma_1'$ with the use of $k=n-1$ rules and $\sigma_1'$ can be transformed
	to $\sigma_2$ with the use of a single rule. By the induction hypothesis,
	we conclude that $\state{M}{H}\fctrans{\sigma_1'}\state{M_1}{H_1'}$, where 
	$H_1\asymp H_1'$. We need to show that $\state{M_1}{H_1'}\asymp 
	\state{M_2}{H_2}$. Let us write
	$\sigma_1' = u;w;v$ and $\sigma_2 = u;w';v$, where $w$, $w'$ refer
	to the parts of the two executions where the equivalence rule has been applied.
	Furthermore, suppose that
	$\state{M}{H}\fctrans{u}\state{M_u}{H_u}\fctrans{w}\state{M_w}{H_w}\fctrans{v}\state{M_1}{H_1'}$ and
	$\state{M}{H}\fctrans{u}\state{M_u}{H_u}\fctrans{w'}\state{M_w'}{H_w'}\fctrans{v}\state{M_2}{H_2}$.
	Three cases exist:
	\begin{enumerate}
		\item $w= t_1;t_2$ and $w'=t_2;t_1$ with $t_1$ and $t_2$ concurrent
		\item $w=t;\underline{t}$ and $w'=\epsilon$
		\item $w=\underline{t};t$ and $w'=\epsilon$
	\end{enumerate}
	In all the cases above, we have that $\state{M_w}{H_w}\asymp \state{M_w'}{H_w'}$:
	for (a) this follows by the definition of concurrent transitions, whereas
	for (b) and (c) by the Loop Lemma. Given the equivalence of these two
	states, by Corollary~\ref{equivalent-executions}, we have that $\state{M_w}{H_w}\fctrans{v}\state{M_1}{H_1'}$  and
	$\state{M_w'}{H_w'}\fctrans{v}\state{M_2}{H_2}$, where $\state{M_1}{H_1'}
	\asymp \state{M_2}{H_2}$, as required. This completes the proof.
	\proofend
\end{enumerate}

\subsection{Out-of-causal-order reversibility}

While in backtracking and causal-order reversibility reversing is cause respecting, there are many examples of systems where undoing
actions in an out-of-causal order is either inherent or
desirable. In this section we consider this type of 
reversibility in the context of RPNs. We begin by specifying that in out-of-causal-order reversibility any executed transition can be reversed at any time.
\begin{definition}\label{o-enabled}{\rm
		Consider a reversing Petri net $(A,P,B,T,F)$, a state $\state{M}{H}$, and a transition $t\in T$. We say that $t$ is \emph{$o$-enabled} in 
		$\state{M}{H}$, if $H(t)\neq \es$.
}\end{definition}

\begin{figure}[t]
	\centering
	\subfigure{\includegraphics[width=5.65cm]{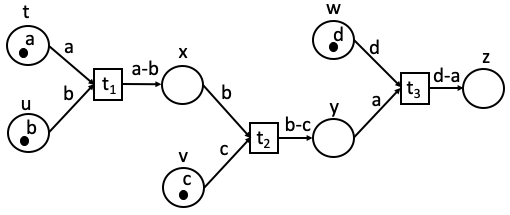}}
	\subfigure{\includegraphics[width=.56cm]{arrow1.png}}
	\subfigure{\includegraphics[width=.54cm]{arrow2.png}}
	\subfigure{\includegraphics[width=.55cm]{arrow3.png}}
	\subfigure{\includegraphics[width=6cm]{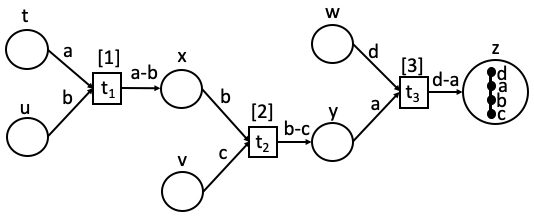}}
	\subfigure{\includegraphics[width=.56cm]{arrow1r.png}}
	\subfigure{\includegraphics[width=6cm]{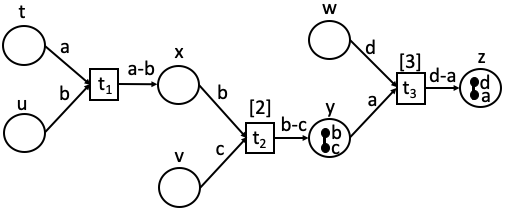}}
	\subfigure{\includegraphics[width=.56cm]{arrow2r.png}}
	\subfigure{\includegraphics[width=6cm]{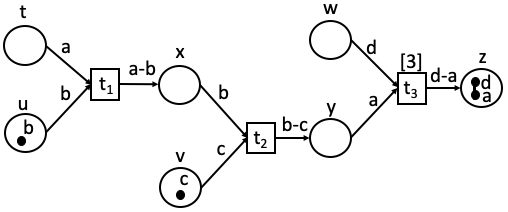}}
	\subfigure{\includegraphics[width=.56cm]{arrow3r.png}}
	\subfigure{\includegraphics[width=6cm]{outoforder6.png}}
	\setlength{\abovecaptionskip}{-1pt}
	\caption{Out-of-causal-order example}\label{o-example}
\end{figure}

Let us begin to consider out-of-causal-order reversibility
via the example
of Figure~\ref{o-example}. The first two nets in the figure present the forward execution of the transition sequence $\langle t_1,t_2,t_3\rangle$. 
Suppose that transition $t_1$ is to be reversed out of order. 
The effect of
this reversal should be the destruction of the bond between $a$ and $b$. This means that the component
$d\bond a\bond b\bond c$ is broken into the bonds $d\bond a$ and $b\bond c$, which should
backtrack within the net to capture the reversal of the transition. Nonetheless, the tokens
of $d\bond a$ must remain at place $z$. This is because a bond exists between them that has not been reversed
and was the effect of the immediately preceding transition $t_3$.
However, in the case of $b\bond c$, the bond can be returned to place $y$, which is the place
where the two tokens were connected and from where they could continue to participate in any further
computation requiring their coalition. Once transition $t_2$ is subsequently reversed,
the bond between $b$ and $c$ is destroyed and thus the two tokens are able to return to their
initial places as shown in the third net in the figure. Finally, when subsequently transition $t_3$ is reversed, the bond
between $d$ and $a$ breaks and, given that neither $d$ nor $a$ are connected to other elements,
the tokens return to their initial places. 
As with the other types of reversibility, when reversing a transition histories are updated by removing the greatest key identifier of the executed transition. 

Summing up, the effect of reversing a transition in out-of-causal order is that all bonds created by the transition
are undone. This may result in tokens backtracking in the net. Further,
if the reversal of a transition causes a coalition of bonds 
to be broken down into
a set of subcomponents due to the destruction of bonds, then each of these coalitions should
flow back, as far back as possible, after the last transition in which this sub-coalition participated. 
To capture this notion of ``as far backwards as possible''
we introduce the following: 

\begin{definition}\label{last}{\rm
		{	Given a reversing Petri net $(A,P,B,T,F)$, an initial marking $M_0$,  
			a history $H$, and 
			a set of bases and bonds $C\subseteq A\cup B$ we write:
			\[
			\begin{array}{rcl}
			\lastt{C,H} &=& \left\{
			\begin{array}{ll}
			t , \;\;\textrm{ if }\exists t, \; \effects{t}\cap C\neq \emptyset, \; H(t)\neq \emptyset, \mbox{ and }\\
			\hspace{0.25in} \not\exists t', \;\effects{t'} \cap C\neq \emptyset, \; H(t') \neq \emptyset \; , 	\\ 	
			\hspace{0.25in} max(H(t'))\geq max (H(t)) \\
			\bot,  \;\textrm{ otherwise }
			\end{array}
			\right.
			\end{array}
			\]
			\[
			\begin{array}{rcl}
			\lastp{C,H} &=& \left\{
			\begin{array}{ll}
			x , \;\;\textrm{ if }t=\lastt{C,H}, \{x\} = \{y\in t\circ 
			\mid F(t,y)\cap C \neq \emptyset\}\\
			\hspace{0.25in} \textrm{or, if } \bot=\lastt{C,H}, C\subseteq M_0(x)\\
			\bot,  \;\textrm{ otherwise }
			\end{array}
			\right.
			\end{array}
			\]}
}\end{definition}
Thus, if component $C$ has been manipulated by some previously-executed
transition,  then $\lastt{C,H}$ is the last executed such transition.
Otherwise, if no such transition exists (e.g., because all transitions
involving $C$ have been reversed), then $\lastt{C,H}$ is undefined 
($\bot$). Similarly, $\lastp{C,H}$ is the outgoing place connected to
$\lastt{C,H}\neq \bot$ having common tokens with $C$, assuming that such
a place is unique, or the place in the initial marking in which $C$ 
existed if $\lastt{C,H}= \bot$, and undefined otherwise.

Transition reversal in an out-of-causal order can thus be defined as follows: 
\begin{definition}\label{oco-def}{\rm
		Given a reversing Petri net $(A, P,B,T,F)$, an initial marking $M_0$, a state $\langle M, H\rangle$ and a transition $t$ that is $o$-enabled in $\state{M}{H}$, we write 
		$\state{M}{H}
		\otrans{t} \state{M'}{H'}$
		where $H'$ is defined as in Definition~\ref{br-def} and  we have:
		\begin{eqnarray*}
			M'(x) & = & \Big(M(x)\cup \bigcup_{a\in M(y){\cap \effects{t}}, \lastp{C_{a,y},H'} =x}C_{a,y}\Big ) \\ 	 
			&& -\Big(\effect{t}  \cup \bigcup_{a\in M(x){\cap \effects{t}}, \lastp{C_{a,x},H'}\neq x}C_{a,x}\Big) 
		\end{eqnarray*}
		where we use the shorthand $C_{b,z} = \connected(b,M(z)-\effect{t})$ for $b\in A$, $z\in P$. 
}\end{definition}

Thus, when a transition $t$ is reversed in an out-of-causal-order 
fashion all bonds that were created by the transition in $\effect{t}$ 
are undone. Furthermore, tokens and bonds involved in the transition are
relocated back to the place where they would have existed if 
transition $t$ never took place, as defined by $\lastp{C,H'}$. Note
that if the destruction of a bond divides a component 
into smaller connected sub-components then each of these
sub-components is relocated separately. 
Specifically, the definition states that: if a token $a$ and 
its connected components involved in transition $t$,
last participated in some transition 
with outgoing place $y$ other than $x$,  then the sub-component
is removed from place $x$ and returned to place $y$, otherwise
it is returned to the place where it occurred in the initial marking. 

\begin{figure}[t]
	\centering
	\subfigure{\includegraphics[width=5.5cm]{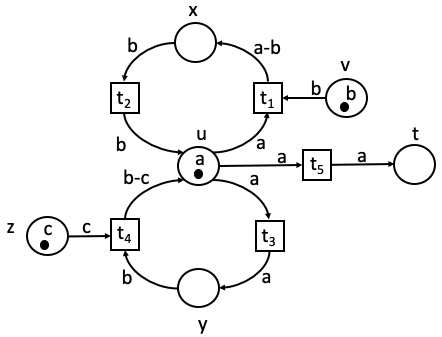}}
	\subfigure{\includegraphics[width=.6cm]{sigmacycle.png}}
	\subfigure{\includegraphics[width=5.5cm]{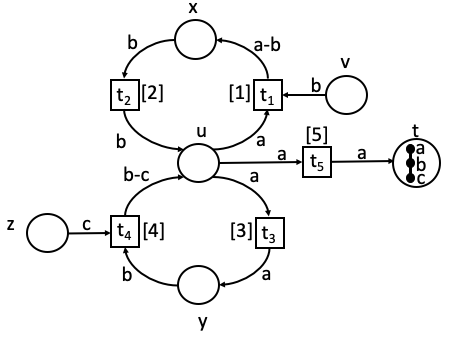}}\\
	\subfigure{\includegraphics[width=.7cm]{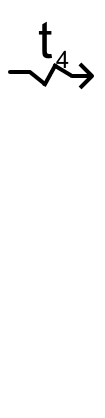}}
	\subfigure{\includegraphics[width=5.5cm]{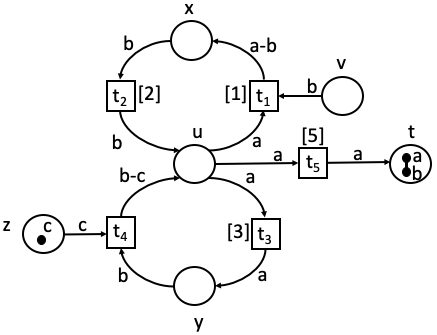}}
	\caption{Out-of-causal-order reversing where $\sigma=\langle t_1,t_2,t_3,t_4,t_5\rangle$.}
	\label{o-cycles}
\end{figure}

An example of out-of-causal-order reversibility in a cyclic RPN can be seen in Figure~\ref{o-cycles}. Here the cycles $\langle t_1,t_2\rangle$ and  $\langle t_3,t_4\rangle$ are executed in this order followed by transition $t_5$. We reverse in out-of-causal order transition $t_4$, which breaks the bond between $b \bond c$ and returns token $c$ back to its original place $z$. Moreover, the bond between $a\bond b$ remains in place $t$, which is the outgoing place of the last transition of token $a$. Note that this state did not occur during the forward execution of the RPN.

The following results describe how tokens and bonds are manipulated
during out-of-causal-order reversibility, where we write $\fotrans{}$ for $\trans{}\cup\otrans{}$.
\begin{proposition}\label{markings}{\rm
		Suppose $\state{M}{H} \fotrans{t}\state{M'}{H'}$ and let $a\in A$ where
		$a\in M(x)$ and $a\in M'(y)$.
		Then, 
		$\connected(a,M'(y))=\connected(a,M(x)\cup C)$, where $C=\effect{t} \cup\{ \connected(b,M(u))\mid a-b\in \effect{t}, b\in M(u)\})$,
		if $t$ is a forward transition, and $\connected(a,M'(y))=\connected(a,M(x)-\effect{t})$, if $t$ is a
		reverse transition.
	}
\end{proposition}
\paragraph{Proof:}
The proof is straightforward by the definition of the firing rules.
\proofend

\begin{proposition}\label{prop5}
	\rm Given a reversing Petri net $(A, P,B,T,F)$, an initial state 
	$\langle M_0, H_0\rangle$, and an execution
	$\state{M_0}{H_0} \fotrans{t_1}\state{M_1}{H_1} \fotrans{t_2}\ldots 
	\fotrans{t_n}\state{M_n}{H_n}$ the following hold for all $0\leq i \leq n$:
	\begin{enumerate}
		\item For all $a\in A$, $|\{x\in P \mid a\in M_i (x)\}| = 1$, 
		and  $a\in M_i(x)$ where 
		$x=\lastp{\connected(a,M_{i}(x)),H_i}$. 
		\item For all $\beta \in B$,  
		\begin {enumerate}
		\item $0 \leq |\{x \in P \mid \beta\in M_i(x)\}| \leq 1$.
		\item if $|\{x \in P \mid \beta\in M_{i-1}(x)\}|=0$ and $|\{x \in P \mid \beta\in M_i(x)\}|=1$,
		then $t_i$ is a forward transition and  $\beta\in \effect{t_i}$,
		\item if $|\{x \in P \mid \beta\in M_{i-1}(x)\}|=1$ and $|\{x \in P \mid \beta\in M_i(x)\}|=0$,
		then $t_i$ is a reverse transition and  $\beta\in \effect{t_i}$,
		\item if $|\{x \in P \mid \beta\in M_{i-1}(x)\}|=|\{x \in P \mid \beta\in M_i(x)\}|$,
		then $\beta\not\in \effect{t_i}$.
	\end{enumerate}
\end{enumerate}
\end{proposition}
\paragraph{Proof:}
Consider a reversing Petri net $(A, P,B,T,F)$, an initial state 
$\langle M_0, H_0\rangle$, and an execution
$\state{M_0}{H_0} \fotrans{t_1}\state{M_1}{H_1} \fotrans{t_2}\ldots \fotrans{t_n}\state{M_n}{H_n}$. The 
proof is by induction on $n$.
\paragraph{Base Case.} For  $n=0$, by our assumption of token 
uniqueness and the definitions of $\mathsf{last}_P$ and
$\mathsf{last}_T$ the claim follows trivially.
\paragraph{Induction Step.} Suppose the claim holds for all but the last transition
and consider
transition $t_n$. Two cases exist, depending on whether $t_n$ is a forward or a reverse transition:
\begin{itemize}
\item Suppose that $t_n$ is a forward transition. Then by Proposition \ref{prop1},  for all $a\in A$,
$|\{x\in P \mid a\in M_n (x)\}| = 1$. 
Additionally, we may see that if $a\in M_n(x)$ two cases exists.
If $a\in \connected(b,M_{n-1}(y))$, for some $b\in F(t_n,z)$
then $x=z=\lastp{\connected(a,M_n(x)), H_n}$. 
Otherwise, it must be that $a\in M_{n-1}(x)$
where, by the induction hypothesis, $x = \lastp{\connected(a,M_{n-1}(x)), H_{n-1}}$. Since $a\not \in \effect{t_n}$, by clause 2(b)  we may deduce that    $\connected(a,M_{n-1}(x))=\connected(a,M_{n}(x))$, which 
leads to 
$x = \lastp{(\connected(a,M_{n-1}(x)), H_{n-1}}=\lastp{\connected(a,M_n(x)), H_n}$. Thus, the result follows.

Now let $\beta \in B$.
To begin with, clause (2)(a) follows by Proposition \ref{prop1}. 
Furthermore, we may see that the forward transition $t_n$ may only create 
exactly the bonds in $\effect{t_n}$ and it maintains all remaining bonds. Thus,
clauses 2(b) and 2(d) follow. 
\item {Suppose that $t_n$ is a reverse transition. 
	Consider $a\in A$ with $a\in M_{n-1}(x)$ for some $x\in P$. Two cases exist:
	\begin{itemize}
		\item Suppose $\lastt{\connected(a,M_{n-1}(x)-\effect{t_n}),H_n} = \bot$. Then, it  must be that $\connected(a,M_{n-1}(x)-\effect{t_n})
		\subseteq M_0(y)$ for some $y$ such that $a\in M_0(y)$. Suppose
		that this is not the case. Then there must exist some $\beta\in\connected(a,M_{n-1}(x)-\effect{t_n})$ with $\beta\not\in M_0(y)$. By the induction hypothesis, there exists some $t_i$
		in the execution such that $\beta\in \effect{t_i}$ which was not
		reversed, i.e. $H_n(t_i)\neq \emptyset$. This however implies
		that $t_i$ is a transition that has manipulated the connected
		component $\connected(a,M_{n-1}(x)-\effect{t_n})$, which contradicts
		our assumption of $\lastt{\connected(a,M_{n-1}(x)-\effect{t_n}),H_n} = \bot$. 
		Therefore, $a\in M_n(y)$, where $a\in M_0(y)$ and by Proposition~\ref{markings} $\connected(a,M_{n-1}(x)-\effect{t_n})=\connected(a,M_{n}(y))$ which gives $y = \lastp{\connected(a,M_n(y)), H_n}$
		and the result follows.
		\item Suppose $\lastt{\connected(a,M_{n-1}(x)-\effect{t_n}),H_n} = t_k$. 
		Then, it  must be that there exists a unique $y\in t_k
		\circ$ 
		such that $\connected(a,M_{n-1}(x)-\effect{t_n})
		\cap F(t_k,z)\neq \emptyset$. Suppose
		that this is not the case. Then there must exist some $\beta=(a,b)\in \connected(a,M_{n-1}(x)-\effect{t_n})$ with $a\in F(t_k,y_1)$, $b\in F(t_k,y_2)$, and $y_1\neq y_2$. Since $\beta\in M_n(y)$,
		by the induction hypothesis, there exists some $t_i$
		in the execution such that $\beta\in \effect{t_i}$, $i>k$ which was not
		reversed, i.e. $H_n(t_i)\neq \emptyset$. This however implies
		that $t_i$ is a transition that has manipulated the connected
		component $\connected(a,M_{n-1}(x)-\effect{t_n})$ later than 
		$t_k$, which contradicts
		our assumption of $\lastt{\connected(a,M_{n-1}(x)-\effect{t_n}),H_n} = t_k$. Therefore, there exists a unique $y\in t_k
		\circ$ 
		such that $\connected(a,M_{n-1}(x)-\effect{t_n})
		\cap F(t_k,z)\neq \emptyset$, $a\in M_n(y)$. Furthermore, by Proposition~\ref{markings} $\connected(a,M_{n-1}(x)-\effect{t_n})=\connected(a,M_{n}(y))$ which gives
		$y = \lastp{\connected(a,M_n(y)), H_n}$
		and the result follows.
	\end{itemize}
	
	Now consider $\beta \in B$. By clause 1, we may deduce clause 2(a). 
	Finally, we may observe that the reverse transition $t_n$ may only remove 
	exactly the bonds in $\effect{t_n}$ and it maintains all remaining bonds, thus,
	clauses 2(b)-2(d) follow. 
}
\proofend
\end{itemize}
As we have already discussed (e.g., see Figures~\ref{catalyst2} and~\ref{o-cycles}), unlike causal-order 
reversibility, out-of-causal-order reversibility may give rise
to states that cannot be reached by forward-only execution.
Nonetheless, note that the proposition establishes
that during out-of-causal-order reversing
it is not the case that tokens and bonds may reach places they
have not previously occurred in. On the contrary,
%
a component will always return to the place following the last
transition that has manipulated it. This observation also gives rise to the following corollary, which 
characterises the marking of a state during computation.
\begin{corollary}\label{oco-tokenplace}{\rm Given a reversing Petri net $(A, P,B,T,F)$, an initial state 
	$\langle M_0, H_0\rangle$, and an execution
	$\state{M_0}{H_0} \fotrans{t_1}\state{M_1}{H_1} \fotrans{t_2}\ldots 
	\fotrans{t_n}\state{M_n}{H_n}$, then for all $x\in P$ we have 
	\[ M_n(x) =\bigcup_{a\in M_n(y),\lastp{C_{a,y},H_n} =x} C_{a,y}
	\]
	where $C_{a,y} = \connected(a,M_n(y))$}.
\end{corollary}
\paragraph{Proof:}
According to Proposition~\ref{prop5} clauses (1) and 2(a) the result follows.
\proofend

The dependence of the position of a connected component and a transition
sequence can be exemplified by the following proposition.
\begin{proposition}\label{second}{\rm Consider executions $\state{M_0}{H_0} 
	\fotrans{\sigma_1} \state{M_1}{H_1}$, $\state{M_0}{H_0} 
	\fotrans{\sigma_2} \state{M_2}{H_2}$,
	and a 
	token $a$ such that $a\in M_1(x)$, $a\in M_2(y)$, for some $x$, $y\in P$, and $\connected(a, M_1(x)) = \connected(a,M_2(x))$. Then,  
	$\lastt{\connected(a,M_1(x)),H_1}=\lastt{\connected(a,M_2(y)),H_2}$ implies
	$x=y$.
}\end{proposition}

\paragraph{Proof:} 
Consider executions $\state{M_0}{H_0} \fotrans{\sigma_1} 
\state{M_1}{H_1}$, $\state{M_0}{H_0} \fotrans{\sigma_2} 
\state{M_2}{H_2}$  and a token $a$ such that $a\in M_1(x)$,
$a\in M_2(x)$.
Further, let us assume that $\lastt{\connected(a,M_1(x)),H_1} = \lastt{\connected(a,M_2(y)),H_2}$. Two cases exist:
\begin{itemize}
\item $\lastt{\connected(a,M_1(x)),H_1} = \lastt{\connected(a,M_2(y)),H_2}=\bot$.
This implies that no transition has manipulated any of the tokens
and bonds of the two connected components. As such, by Proposition~\ref{prop5}, $\connected(a,M_1(x))\subseteq M_0(x)$ and $\connected(a,M_2(y))\subseteq M_0(y)$,
and by the uniqueness of tokens we conclude that $x=y$ as required.
\item   $\lastt{\connected(a,M_1(x)),H_1} = \lastt{\connected(a,M_2(y)),H_2} = t$. 
This implies that there is $b\in \connected(a,M_1(x))=\connected(a,M_2(y))$ such that $b\in F(t,z)$ for some place $z$.
By definition, we deduce that $ \lastp{\connected(a,M_1(x)),H_1}$ $=z=\lastp{\connected(a,M_2(y)),H_2}$, thus, $x=y$ as required.
\proofend
\end{itemize}

From the above result we may prove the following proposition establishing that 
executing two causally equivalent sequences of transitions
in the out-of-causal-order setting will give rise to causally equivalent states.

\begin{proposition}\label{corTheorem1}{\rm\ \  
	Suppose 
	$\state{M_0}{H_0} \fotrans{\sigma_1} \state{M_1}{H_1}$ and $\state{M_0}{H_0} \fotrans{\sigma_2} \state{M_2}{H_2}$. 
	If $\sigma_1\asymp\sigma_2$ then  $\state{M_1}{H_1}\asymp\state{M_2}{H_2}$.
}
\end{proposition} 
\paragraph{Proof:} Suppose $\state{M_0}{H_0} \fotrans{\sigma_1} \state{M_1}{H_1}$,
$\state{M_0}{H_0} \fotrans{\sigma_2} \state{M_2}{H_2}$ and $\sigma_1\asymp\sigma_2$. 
Since $\sigma_1\asymp \sigma_2$ it must be that the two executions
contain the same causal paths, therefore,
$H_1\asymp H_2$. To show that $M_1 = M_2$ consider
token $a$ such that $a\in M_1(x)\cap M_2(y)$. 
Since $\sigma_1\asymp \sigma_2$, we may conclude that the two executions contain the same set of executed and not reversed transitions. Thus, by Proposition~\ref{prop5}(2), 
we have $\connected(a, M_1(x)) = \connected(a, M_2(y))$. Furthermore,
it must be that $t_1=\lastt{\connected(a, M_1(x)), H_1}
=\lastt{\connected(a, M_2(y)), H_2}= t_2$. 
If not, since $\sigma_1 \asymp \sigma_2$, we would have that $t_1$
and $t_2$ are concurrent, which is not possible since they
manipulate the same connected component and thus a
causal relation exists between them. Therefore, by Proposition~\ref{second},
$x=y$. This implies by Corollary~\ref{oco-tokenplace}
that $M_1(x) = M_2(x)$, for all places $x$, which completes the proof.
\proofend

We finally establish a Loop Lemma for out-of-causal reversibility.
\begin{lemma}[Loop]\label{loopo}{\rm 
	For any forward transition $\state{M}{H}\trans{t}\state{M'}{H'}$ there exists a reverse
	transition $\state{M'}{H'} \otrans{t} \state{M}{H}$. 
}\end{lemma}
\paragraph{Proof:}
Suppose $\state{M}{H}\trans{t}\state{M'}{H'}$. Then $t$ is clearly $o$-enabled 
in $H'$. Furthermore, $\state{M'}{H'} \otrans{t} \state{M''}{H''}$ where $H''=H$
by the definition of $\otrans{}$. 
In addition, for all $a\in A$, we may prove that $a\in M''(x)$ if and only if $a\in M(x)$. Suppose $a\in M(y)$, we distinguish two cases. 
If $\connected(a,M(y))\cap \guard{t}= \es$, then we may see that $a\in M'(y)$ and 
$a\in M''(y)$, and the result follows. Otherwise, if $\connected(a,M(y))\cap \guard{t}\neq \es$, then 
$a\in M'(z)$, where $F(t,z)\cap \connected(a,M(y))\neq \es$.
Furthermore, suppose that $a\in M''(w)$. By
Proposition~\ref{markings} we have that $\connected(a,M'(z))=\connected(a,M(y)\cup C)$, $C=\effect{t}\cup\{ \connected(b,M(u))\mid a-b\in \effect{t}, b\in M(u)\}$, and 
$\connected(a,M''(w))= \connected(a,M'(z)- \effect{t})
=\connected(a,(M(y)\cup C)-\effect{t}) 
=\connected(a,M(y))$. 
Furthermore, $y = \lastp{\connected(a,M(y)), H}$, by Corollary~\ref{oco-tokenplace}. Since $H= H''$, we
have $w=\lastp{\connected(a,M''(w)), H'')}= \lastp{\connected(a,M(y)), H)} = y$, and the result follows.
\proofend

Note that in the case of out-of-causal-order reversibility, the opposite direction of the 
lemma does not hold. This is because reversing a transition in an out-of-causal-order
fashion may bring a system to a state not reachable by forward-only transitions, and
where the transition is not enabled in the forward direction. As an example, 
consider the RPN of Figure~\ref{o-example} and after the reversal of transition 
$t_2$. In this state, transition $t_2$ is not forward enabled since token $b$ is
not available in place $x$, as required for the transition to fire.

\subsection{Relationship between reversibility notions}
We continue to study the relationship between the three forms of reversibility. Our first result confirms
the relationship between the enabledness conditions for each of backtracking, causal-order, and out-of-causal-order reversibility.
\begin{proposition}\label{enable}{\rm\ \
	Consider  a state $\langle M, H\rangle$, and a transition $t$. Then, if $t$ is $bt$-enabled in  $\langle M, H\rangle$
	it is also $co$-enabled. Furthermore, if  $t$ is $co$-enabled in  $\langle M, H\rangle$
	then it is also $o$-enabled.
}
\end{proposition}
\paragraph{Proof:}
The proof is immediate by the respective definitions.
\proofend

We next demonstrate a ``universality'' result of the $\otrans{}$ transition relation by showing that it manipulates the
state of a reversing Petri net in an identical way to $\ctrans{}$, in the case of $co$-enabled transitions,
and to $\btrans{}$, in the case of $bt$-enabled transitions.
Central to the proof is the following result establishing
that during causal-order reversibility a component is
returned to the place following the last transition
that has manipulated it or, if no such transition exists,
in the place where it occurred in the initial marking. 
\begin{proposition}\label{last-as-co}
\rm Given a reversing Petri net $(A, P,B,T,F)$, an initial state 
$\langle M_0, H_0\rangle$, and an execution
$\state{M_0}{H_0} \fctrans{t_1}\state{M_1}{H_1} \fctrans{t_2}\ldots 
\fctrans{t_n}\state{M_n}{H_n}$. Then for all $a\in A$, 
$a\in M_n(x)$ where $x=\lastp{\connected(a,M_{n}(x)),H_n}$.
\end{proposition}
\paragraph{Proof:}
The 
proof is by induction on $n$ and it follows along similar lines to the proof of Proposition~\ref{prop5}(1).
\proofend

Propositions~\ref{prop5} and~\ref{last-as-co} yield the following
corollary for forward-only execution.
\begin{corollary}\label{last-as-f}
\rm Given a reversing Petri net $(A, P,B,T,F)$, an initial state 
$\langle M_0, H_0\rangle$, and an execution
$\state{M_0}{H_0} \trans{t_1}\state{M_1}{H_1} \trans{t_2}\ldots 
\trans{t_n}\state{M_n}{H_n}$, for all $a\in A$, 
$a\in M_n(x)$ where $x=\lastp{\connected(a,M_{n}(x)),H_n}$.
\end{corollary}
We may now verify that the causal-order and out-of-causal-order reversibility have the same effect when reversing a $co$-enabled transition.
\begin{proposition}\label{c-to-o}{\rm\ \
	Consider a state $\langle M, H\rangle$ and a transition $t$ $co$-enabled in  $\langle M, H\rangle$.
	Then, $\langle M, H\rangle\ctrans{t}\langle M', H'\rangle$ if and only if $\langle M, H\rangle\otrans{t}\langle M', H'\rangle$.
}
\end{proposition}
\paragraph{Proof:}
Let us suppose that transition $t$ is $co$-enabled and $\langle M, H\rangle\ctrans{t}
\langle M_1, H_1\rangle$. 
By Proposition~\ref{enable}, $t$ is also $o$-enabled. Suppose $\langle M, H\rangle\otrans{t}\langle M_2, H_2\rangle$.
It is easy to see that in fact $H_1 = H_2$ (the two histories are 
as $H$ with the exception that $H_1(t) = H_2(t) = H(t)-\{\max(H(t))\}$). 

To show that $M_1=M_2$ first we observe that for all
$a\in A$, by Proposition~\ref{last-as-co} we have $a\in M_1(x)$ where
$x=\lastp{\connected(a,M_1(x)),H_1}$ and by
Proposition~\ref{prop5} we have $a\in M_1(y)$ where
$y=\lastp{\connected(a,M_2(y)),H_2}$. We may also see
that $\connected(a,M_1(x)) = \connected(a,M(z)-\effect{t})= \connected(a,M_2(y))$, where $a\in M(z)$. Since in addition we have $H_1=H_2$
the result follows.

Now let $\beta\in B$. We must show that
$\beta\in M_1(x)$ if and only if $\beta\in M_2(x)$. Two cases exist:
\begin{itemize}
\item If $\beta \in \effect{t}$ then by Propositions~\ref{Prop4} and~\ref{prop5}, $\beta\not\in M_1(x)$ and $\beta\not \in M_2(x)$ for all $x\in P$.
\item if $\beta \not\in \effect{t}$ then by Propositions~\ref{Prop4} and~\ref{prop5},
$|\{x\in P \mid \beta\in M_1(x))\}| = |\{x\in P \mid \beta\in M_2(x))\}|= 1$
and by the analysis on
tokens  $\beta\in M_1(x)$ if and only if $\beta\in M_2(x)$ and
the result follows.
\end{itemize}
This completes the proof.
\proofend

An equivalent result can be obtained for backtracking.

\begin{proposition}\label{b-to-c}{\rm\ \
	Consider  a state $\langle M, H\rangle$, and a transition $t$, $bt$-enabled in  $\langle M, H\rangle$.
	Then, $\langle M, H\rangle\btrans{t}\langle M', H'\rangle$ if and only if $\langle M, H\rangle\otrans{t}\langle M', H'\rangle$.
}
\end{proposition}
\paragraph{Proof:}
Consider a state $\langle M, H\rangle$ and suppose
that transition $t$ is $bt$-enabled and $\langle M, H\rangle\btrans{t} \langle M', H'\rangle$. Then,
by Proposition~\ref{enable}{, there exists $k\in H(t)$, such that {for all $t'\in T$, $k'\in H(t')$, it holds that
	$k\geq k'$}. This implies
that} $t$ is also $co$-enabled, and by the definition of $\ctrans{}$, we conclude that 
$\langle M, H\rangle{\ctrans{t}} \langle M', H'\rangle$. Furthermore, by Proposition~\ref{c-to-o}
$\langle M, H\rangle{\otrans{t}} \langle M', H'\rangle$, and the result follows.
\proofend

We obtain the following corollary confirming
the expectation that backtracking is an instance of causal reversing, which in turn is an instance of
out-of-causal-order reversing.  It is easy to see that both inclusions are strict,
as for example illustrated in Figures~\ref{b-example},~\ref{co-example}, and~\ref{o-example}.
\begin{corollary}\label{connect}{\rm\ \
	$\btrans{} \subset\ctrans{}\subset \otrans{}$. 
}
\end{corollary}

\paragraph{Proof:}
The proof follows from Propositions~\ref{c-to-o} and~\ref{b-to-c}.
\proofend

We note that in addition to establishing the relationship between the three notions of reversibility,
the above results provide a unification of the different reversal strategies, in the sense that
a single firing rule, $\otrans{}$, may be paired with the three notions of transition enabledness
to provide the three different notions of reversibility.
This fact may be exploited in the proofs of results that span the three notions of reversibility. Such a proof follows in
the following proposition that
establishes a reverse diamond property
for RPNs. According to this property, the execution of a reverse
transition does not preclude the execution of another
reverse transition and their execution leads to the same
state.  In what follows we write $\frtrans{}$ 
for $\trans{}\cup\rtrans{}$ where $\rtrans{}$ could be an instance of one of $\btrans{}$, $\ctrans{}$, and $\otrans{}$.  

\begin{proposition}[Reverse Diamond]\label{reverseDiamond}{\rm 
	Consider a state $\state{M}{H}$, and reverse transitions $\state{ M}{H} \rtrans{t_1}\state{ M_1}{H_1}$ and $ \state{ M}{H} \rtrans{t_2}\state{ M_2}{H_2}$, $t_1\neq t_2$. Then $\state{ M_1}{H_1} \rtrans{t_2}\state{ M'}{H'}$ and $\state{ M_2}{H_2} \rtrans{t_1}\state{ M'}{H'}.$
}\end{proposition}

\paragraph{Proof:}
Let us suppose that
$\state{ M}{H} \rtrans{t_1}\state{ M_1}{H_1}$ and $ \state{ M}{H} \rtrans{t_2}\state{ M_2}{H_2}$, $t_1\neq t_2$.
First we note that $\rtrans{}$ may be an instance of $\ctrans{}$ or $\otrans{}$ but not $\btrans{}$, since
in the case of $\btrans{}$ the backward transition is uniquely determined
as the transition with the maximum key. 
Furthermore, we observe that 
$t_2$ remains backward-enabled in $\state{ M_1}{H_1}$ and likewise $t_1$ in
$\state{ M_2}{H_2}$. Specifically, if $\rtrans{}=\ctrans{}$,
since $t_1$ and $t_2$  are $co-$enabled in $\state{M}{H}$, by Definition~\ref{co-enabled} we conclude that
$(t_2,max(H(t_2)))$ is not causally dependent on $(t_1,max(H(t_1)))$ and vice versa, which continues
to hold after the reversal of each of these transitions.
In the case of $\rtrans{}=\otrans{}$ this is straightforward from the definition of $o$-enabledness.

So, let us suppose that $\state{M_1}{H_1} \ctrans{t_2}\state{ M_1'}{H_1'}$ and $\state{M_2}{H_2} \ctrans{t_1}\state{ M_2'}{H_2'}$. It is easy to see that $H_1' = H_2'$ since both of 
these histories are identical with $H$ with the maximum keys
of $t_1$ and $t_2$ removed. 

To show that $M_1'=M_2'$ first we observe that for all
$a\in A$, by Propositions~\ref{prop5} and~\ref{last-as-co}
we have $a\in M_1'(x)$, $a\in M_2'(y)$ where
$x=\lastp{\connected(a,M_1'(x)),$ $H_1'}$ and $y=\lastp{\connected(a,M_2'(y)),H_2'}$.
We may see
that $\connected(a,M_1'(x)) = \connected(a,M(z)-(\effect{t_1}\cup \effect{t_2})
= \connected(a,M_2'(y))$, where $a\in M(z)$. Since in addition we have $H_1'=H_2'$
the result follows.

Now let $\beta\in B$. We must show that
$\beta\in M_1'(x)$ if and only if $\beta\in M_2'(x)$. Two cases exist:
\begin{itemize}
\item If $\beta \in \effect{t_1}\cup\effect{t_2}$ then by Propositions~\ref{Prop4} and~\ref{prop5}, $\beta\not\in M_1'(x)$ and $\beta\not \in M_2'(x)$ for all $x\in P$.
\item if $\beta \not\in \effect{t_1}\cup \effect{t_2}$ then by Propositions~\ref{Prop4} and~\ref{prop5},
$|\{x\in P \mid \beta\in M_1'(x))\}| = |\{x\in P \mid \beta\in M_2'(x))\}|$
and, by the analysis on
tokens, $\beta\in M_1'(x)$ if and only if $\beta\in M_2'(x)$.
\end{itemize}
This completes the proof.
\proofend

\begin{corollary}\label{diamondCorollary}{\rm \
	Consider a state  $\state{ M}{H}$, and traces $\sigma_1,\sigma_2$ permutations of the same reverse transitions where $\state{M}{H} \frtrans{\sigma_1}\state{ M'}{H'}$ and  $\state{M}{H} \frtrans{\sigma_2}\state{ M''}{H''}$. Then  $\state{M'}{H'}=\state{M''}{H''}$.
}\end{corollary}

\paragraph{Proof:}
The proof follows by induction on the sum of the length $|\sigma_1|= |\sigma_2|$ and the depth of the earliest 
disagreement between the two traces, and uses similar arguments to those
found in the proof of Proposition~\ref{reverseDiamond}.
\proofend

We note that the analogue of Proposition~\ref{reverseDiamond} for forward
transitions, i.e. the Forward Diamond property, does not hold for RPNs.
To begin with $t_1$ and $t_2$ may be in conflict. 
The proposition fails to hold even in
the case of joinable transitions (i.e. transitions that may yield the same
marking after a sequence of forward moves) due to the case of
co-initial, independent cycles: Even though such cycles can be executed in any order, it is
impossible to complete the square for their initial transitions.

\section{Case Study}\label{sec: CaseStudy}

Biochemical systems, such as covalent bonds, constitute the ideal setting to study reversible computation especially in its out-of-causal-order form. In particular, the \emph{ERK} signalling pathway is one 
of many real-life examples that naturally feature reversibility that violates the causal 
ordering established by forward execution. This pathway has been 
modelled in various formalisms including  \emph{CCSK}~\cite{ERK},  PEPA~\cite{PEPA}, 
BioNetGen~\cite{BioNetGen}, and Kappa~\cite{Kappa}. In this section we illustrate how reversing 
Petri nets allow us to capture naturally this form of out-of-causal-order reversible system.

\begin{figure}
	\centering
	\subfigure{\includegraphics[width=1.8cm]{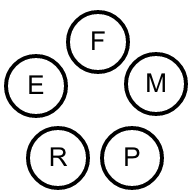}}
	\subfigure{\includegraphics[width=.55cm]{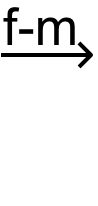}}
	\subfigure{\includegraphics[width=2cm]{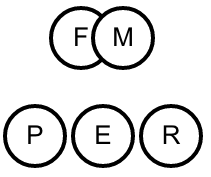}}
	\subfigure{\includegraphics[width=.55cm]{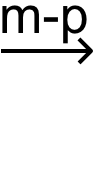}}
	\subfigure{\includegraphics[width=1.6cm]{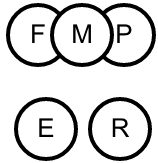}}
	\subfigure{\includegraphics[width=.6cm]{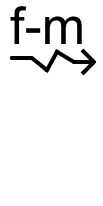}}
	\subfigure{\includegraphics[width=2cm]{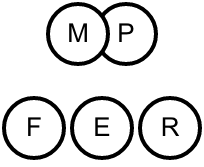}}
	\subfigure{\includegraphics[width=.6cm]{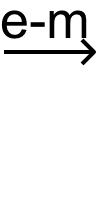}}
	\subfigure{\includegraphics[width=1.6cm]{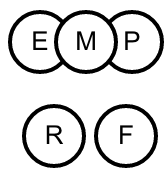}}
	\subfigure{\includegraphics[width=.6cm]{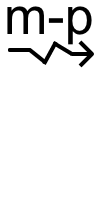}}\\
	\vspace{.5cm}
	\subfigure{\includegraphics[width=2cm]{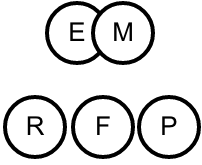}}
	\subfigure{\includegraphics[width=.6cm]{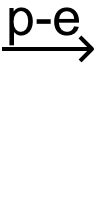}}
	\subfigure{\includegraphics[width=1.6cm]{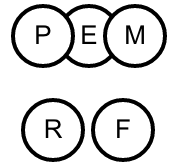}}
	\subfigure{\includegraphics[width=.6cm]{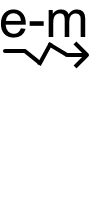}}
	\subfigure{\includegraphics[width=2cm]{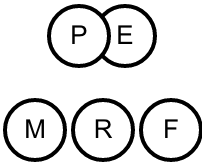}}
	\subfigure{\includegraphics[width=.6cm]{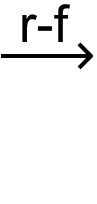}}
	\subfigure{\includegraphics[width=2cm]{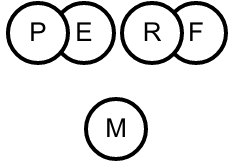}}
	\subfigure{\includegraphics[width=.6cm]{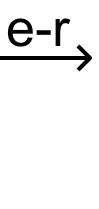}}
	\subfigure{\includegraphics[width=2cm]{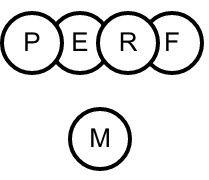}}\\
	\vspace{.5cm}
	\subfigure{\includegraphics[width=.6cm]{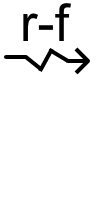}}
	\subfigure{\includegraphics[width=1.6cm]{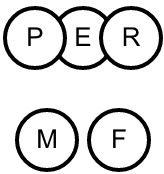}}
	\subfigure{\includegraphics[width=.6cm]{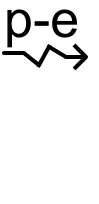}}
	\subfigure{\includegraphics[width=2cm]{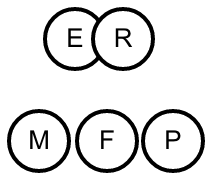}}
	\subfigure{\includegraphics[width=.6cm]{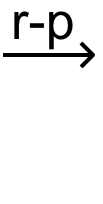}}
	\subfigure{\includegraphics[width=1.6cm]{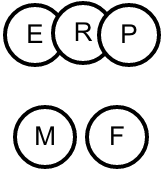}}
	\subfigure{\includegraphics[width=.6cm]{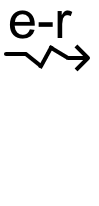}}
	\subfigure{\includegraphics[width=2cm]{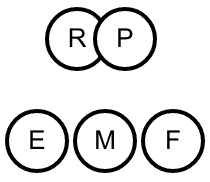}}
	\subfigure{\includegraphics[width=.6cm]{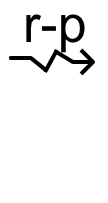}}
	\subfigure{\includegraphics[width=2cm]{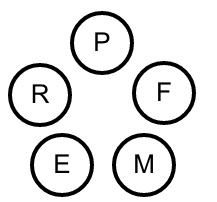}}
	\caption{Reactions in the \emph{ERK}-pathway where \emph{F} denotes \emph{Raf*-1}, 
		\emph{M} denotes \emph{MEK}, \emph{E} denotes \emph{ERK}, \emph{R} denotes \emph{RKIP}, and \emph{P} denotes the phosphorylation of the bonded molecule. }\label{ERK}
\end{figure}

In Figure~\ref{ERK} we demonstrate the extracellular-signal-regulated kinase  (\emph{ERK}) pathway, also known as \emph{Ras/Raf-1, MEK, ERK} pathway, which  is one of the major signalling cassettes of the mitogen activated protein kinase (\emph{EMAPK}) signalling pathway. The ERK pathway is a chain of 
proteins in the cell that delivers mitogenic and differentiation signals from the membrane of a cell 
to the \emph{DNA} in the nucleus, and is regulated by the protein \emph{RKIP}. 
The starting point 
of the pathway is when a signalling molecule binds to a receptor on the cell surface and is spatially organised so that, when a signal arrives at the membrane, it can be transmitted to the nucleus via a cascade of biological reactions that involves protein kinases. A kinase is an enzyme that catalyses the transfer of a phosphate group from a donor molecule to an acceptor. The main $\mathit{MAPK/ERK}$  kinase kinase (\emph{MEKK}) component is the kinase component \emph{Raf-1} that phosphorylates the serine residue on the \emph{MAPK/ERK} kinase \emph{MEK}. We denote  \emph{Raf*-1} with  \emph{F}, \emph{MEK} with  \emph{M}, \emph{ERK} with \emph{E}, \emph{RKIP} with \emph{R}, and  the phosphorylation of the bonded molecule is denoted by {P}.

The pathway begins with the activation of the protein kinase of \emph{Raf-1} by the $G$ protein 
\emph{Ras}  that has been activated near a receptor on the cell's membrane. 
\emph{Ras}  activates a kinase \emph{Raf-1} 
to become \emph{Raf*-1}, which is generally known as a mitogen-activated protein 
kinase kinase kinase (\emph{MAP-KKK}) and can be inhibited by \emph{RKIP}. Subsequently, as we may see in Figure~\ref{ERK}, 
\emph{Raf*-1}~($F$) may bind with \emph{MEK}~($F\bond M$) by facilitating 
the next step in the cascade (\emph{MAPKK}), which is the phosphorylation of the \emph{MEK}~($F \bond M \bond P$) 
protein and the release of \emph{Raf*-1}~($M \bond P$).  
The phosphorylated \emph{MEK}~($M\bond P$) activates a mitogen-activated protein kinase, \emph{ERK}~($E \bond M \bond P$), which in turn becomes 
phosphorylated and releases \emph{MEK}~($P \bond E$). Finally, the  phosphorylation of \emph{MAPK}
allows the phosphorylated \emph{ERK}~($P\bond E$)  to function as an enzyme and translocate in order to signal the nucleus. Now the regulation sequence consumes the phosphorylated \emph{ERK}~($P \bond E$) in order to deactivate  \emph{RKIP}~($R$) from regulating \emph{Raf*-1}~($F$). 
Therefore, when \emph{RKIP} binds \emph{Raf*-1}~($R \bond F$), the resulting complex binds to a 
phosphorylated \emph{ERK}~($P\bond E\bond R\bond F$). In the end, the complex breaks releasing \emph{Raf*-1}~($P \bond E \bond R$), 
which can get involved in the cascade and after the phosphorylation of \emph{RKIP}~($E\bond R\bond P$)  the system releases \emph{ERK}~($R \bond P$) and the phosphorylated  \emph{RKIP}.

\begin{figure}
	\centering
	\setlength{\abovecaptionskip}{-2pt}
	\setlength{\belowcaptionskip}{-2pt}
	\subfigure{\includegraphics[width=13cm]{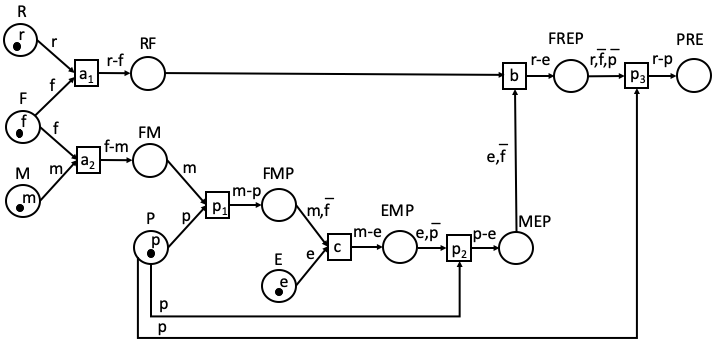}}
	\caption{ERK-pathway example in Reversing Petri Nets}\label{ERK reversing}
\end{figure}

We now describe the biochemical reactions of the ERK signalling pathway as the RPN demonstrated in Figure~\ref{ERK reversing}. On this RPN we represent molecules as tokens that can bond with each other, thus creating more complex molecules, and these composite molecules can be dissolved back to single tokens. The building blocks of the system are the base tokens representing the associated molecules. 

We begin our execution from the already activated  \emph{Raf-1} 
kinase that has become \emph{Raf*-1}. The molecule of \emph{Raf*-1} is represented by base token $f$ and resides in place  $F$. The token availability of base $m$, which represents \emph{MEK} in place $M$ enables the firing of  transition $a_2$ denoting that $f$ has bonded with $m$ and thus creating molecule $f\bond m$. The firing of transition $a_2$ facilitates 
the next step in the cascade, which is transition $p_1$ representing the phosphorylation of $m$ as the binding between $p$ and $m$. Since transition $a_2$ has enabled the execution of $p_1$ the transition $a_2$ is now reversed and therefore releases $f$ back to place $F$. This reversal in necessary
for the next step of the execution where the absence of $f$
is a condition for transition $c$ to fire. Indeed, in transition
$c$, the phosphorylated $m$ is now able to 
activate the kinase \emph{ERK}  denoted by base $e$ and thus creating a bond between $m\bond e$ along with $p$ that shows that $m$ is already phosphorylated. In the next step, transition $p_1$ reverses in order to release $p$, which can then be used in the firing of transition $p_2$ to phosphorylate $e$  and therefore creating the molecule $p\bond e$.  After transition $p_2$ transition $c$ is reversed in order to release $m$ back to $M$. Finally, after the  phosphorylation of $p\bond e$, transition $a_1$ executes in order to bond $f\bond r$ where $r$ represents molecule $RKIP$. Base $r$ functions as an enzyme and  by enabling transition $b$ represents the passing of the signal to the nucleus which can then  consume $p\bond e$ by creating a connected component between $f \bond r \bond e \bond p$. In the end, the complex breaks by reversing $a_1$ in order to release $f$ and $p_2$ to release $p$ which then in action $p_3$ phosphorylates $r$. Finally, the system reverses $b$ to release $e$ followed by the reversal of $p_3$, which releases both $r$ and $p$ and therefore returns the system back to its initial marking.

We show below an execution of the reversing Petri net that illustrates the process until the signal that arrived at the membrane is transmitted  to the nucleus. The following states of the net (with histories omitted) represent a cascade of reactions that involve protein kinases $F$, $M$, $E$, $P$, $R$, with initial marking 
$M_0$ such that  $M_0(R)=\{r\}$, $M_0(F)=\{f\}$,  $M_0(M)=\{m\}$,  $M_0(P)=\{p\}$,  $ M_0(E)=\{e\}$,
and $M_0(p)= \emptyset$ for all remaining places. (In the following, the markings of places with no
tokens are omitted.) 
\begin{tabbing}
	$M_0\xrightarrow{a_2}M_1$,\; \;\= where\; \;\= 
	$M_1(R)=\{r\}$,  $M_1(P)=\{p\}$,  $M_1(E)=\{e\} $,
	\\ \hspace{1.45in} $M_1(FM)= \{f\bond m\} $
	\\
	$M_1\xrightarrow{p_1}M_2$,\> where\> 
	$M_2(R)=\{r\}$,  $  M_2(E)=\{e\} $,
	\\ \hspace{1.45in}  $M_2(FMP)=\{f\bond m, m\bond p\} $
	\\
	$M_2\rtrans{a_2}M_3$,\>  where\> 
	$M_3(R)=\{r\}$,  $M_3(F)=\{f\}$,  $M_3(E)=\{e\} $,
	\\ \hspace{1.45in} $M_3(FMP)=\{m\bond p\} $
	\\
	$M_3\xrightarrow{c}M_4$,\>  where\> 
	$M_4(R)=\{r\}$,  $M_4(F)=\{f\} $,
	\\ \hspace{1.45in} $M_4(EMP)=\{m\bond e, m\bond p\} $  
	\\
	$M_4\rtrans{p_1}M_5$,\>  where\> 
	$M_5(R)=\{r\}$,  $M_5(F)=\{f\} $,  $M_5(P)=\{p\} $,
	\\ \hspace{1.45in} $M_5(EMP)=\{m \bond e\} $
	\\
	$M_5\xrightarrow{p_2}M_6$,\>  where\> 
	$M_6(R)=\{r\}$,  $M_6(F)=\{f\} $,
	\\ \hspace{1.45in} $M_6(MEP)=\{m\bond e, e \bond p\} $  
	\\
	$M_6\rtrans{c}M_7$,\>  where\> 
	$M_7(R)=\{r\}$,  $M_7(F)= \{f\}$,  $M_7(M)=\{m\} $,
	\\ \hspace{1.45in} $M_7(MEP)=\{e\bond p\} $  
	\\
	$M_7\xrightarrow{a_1}M_8$,\>  where\> 
	$M_8(M)=\{m\} $,
	$M_8(RF)=\{r\bond f\} $,    
	\\ \hspace{1.45in} $M_8(MEP)=\{e\bond p\} $ 
	\\
	$M_8\xrightarrow{b}M_9$,\>  where\> 
	$M_9(M)=\{m\} $,  
	$M_9(FREP)=\{r\bond f, r\bond e, e\bond p\} $        
	\\
	$M_9\rtrans{a_1}M_{10}$,\>  where\> 
	$M_{10}(M)=\{m\}$,  $M_{10}(F)=\{f\} $,  
	\\ \hspace{1.45in} $M_{10}(FREP)=\{r\bond e, e\bond p\} $       
	\\
	$M_{10}\rtrans{p_2}M_{11}$,\>  where\> 
	$M_{11}(M) =\{m\}$,  $M_{11}(F)=\{f\}, M_{11}(P)=\{p\} $,  
	\\ \hspace{1.45in}  $M_{11}(FREP)=\{r\bond e\} $    
	\\
	$M_{11}\xrightarrow{p_3}M_{12}$,\>  where\> 
	$M_{12}(M)=\{m\}$,  $  M_{12}(F)=\{f\} $,  
	\\ \hspace{1.45in} $M_{12}(PRE)=\{r\bond e, p\bond r\} $     
	\\
	$M_{12}\rtrans{b}M_{13}$,\>  where\> 
	$M_{13}(M)=\{m\}$,  $ M_{13}(F)=\{f\} M_{13}(E)\{e\} $,  
	\\ \hspace{1.45in} $M_{13}(PRE)=\{p\bond r\} $   
	\\
	$M_{13}\rtrans{p3}M_{14}$,\>  where\> 
	$M_{14}(R)=\{r\}$,  $ M_{14}(F)= \{f\}$,  $ M_{14}(M)=\{m\} $,  \\ \hspace{1.45in} $M_{14}(P)=\{p\}$,  $ M_{14}(E)=\{e\} $
\end{tabbing}


\section{Conclusions}\label{sec:Conclusions}

This paper proposes a reversible approach to Petri nets that allows the modelling
of reversibility as realised by backtracking, causal-order reversing and out-of-causal-order
reversing. To the best of our knowledge, this is the first such proposal in the 
context of Petri nets. For instance, the work of~\cite{PetriNets,BoundedPNs} 
introduces reversed transitions in a Petri net and studies various decidability
problems in this setting. This approach, however, does not precisely capture
reversible behavior due to the property of backward conflict in PNs. Moreover,
the work of~\cite{RPT} proposes a causal semantics for P/T nets by identifying
the causalities and conflicts of a P/T net through unfolding it into an
equivalent occurrence net and subsequently introducing appropriate reverse
transitions to create a coloured Petri net that captures a causal-consistent
reversible semantics. The colours in this net capture causal histories.
On the other hand, our proposal consists of a reversible approach to Petri 
nets, where the formalism supports the reversible semantics without explicitly 
introducing reverse transitions. This is achieved with the use of bonds of 
tokens, which can be thought of as
colours and, combined with the history function of the semantics, capture 
the memory of an execution as needed to implement reversibility. Furthermore, 
the approach allows to implement both causal-order and out-of-causal-order reversibility.

A main main contribution of this work has been the addition of cycles in the RPN model
of~\cite{RPNs}. To enable this, additional machinery has been necessary
to capture causal dependencies in the presence of cycles. As also in~\cite{RPT}, 
our goal has been to allow a causally-consistent semantics reflecting causal 
dependencies as a partial order, and allowing an event to be reversed only if all 
its consequences have already been undone. To achieve this goal we have defined a causal dependence relation that resorts to the \emph{marking} of a net. 
As illustrated via examples (e.g. see Figures~\ref{cycles1} and~\ref{cycles2}), 
this is central in capturing causal dependencies and the intended causal-consistent semantics.

In a related line of work, we are also investigating the expressiveness relationship
between RPNs and CPNs. Specifically, in~\cite{RPNtoCPN} a subclass of RPNs with
trans-acyclic structures has been encoded in coloured PNs. Currently, we are 
extending this work with ultimate objective
to provide and prove the correctness of the translation between the 
two formalisms and analyse the
associated trade-offs in terms of Petri net size. 


Controlling reversibility in RPNs is another topic of current work. Specifically, in~\cite{RC19,ieee}  we have extended RPNs with conditions that control reversibility, and we have applied our framework in the context of wireless communications. This experience has illustrated that resource management can be studied and understood in terms of RPNs as, along with their visual
nature, they offer a number of features, such as token persistence, that is especially relevant in these contexts. In future work, we would like
to further apply our framework in the specific fields as well as in the fields of biochemistry and long-running transactions. 

Finally, our current work focuses on relaxing the remaining restrictions of RPNS,  
to allow multiple tokens of the same base/type to occur in a model and to develop reversible semantics in the presence of bond destruction. %

\paragraph{Acknowledgents:} This research was partially supported by the EU COST Action IC1405.



\bibliographystyle{abbrv}

\end{document}